\documentclass[aps,twocolumn,pre,showpacs,preprintnumbers,eqsecnum,amssymb]{revtex4}

\usepackage{graphicx}
\usepackage{dcolumn}
\usepackage{bm}
\usepackage{amsmath}
\usepackage[T1]{fontenc}

\begin{document}

\preprint{arXiv-extendNLF-v1.tex}

\title{Extended nonlinear feedback model for describing episodes
of high inflation}

\author{Martin A. Szybisz}

\affiliation{Departamento de Econom\'{\i}a, Facultad
de Ciencias Econ\'omicas, Universidad de Buenos Aires,\\
Av. C\'ordoba 2122, RA--1120 Buenos Aires, Argentina}

\author{Leszek Szybisz}
 \altaffiliation{Corresponding author}
 \email{szybisz@tandar.cnea.gov.ar}
\affiliation{Laboratorio TANDAR, Departamento de F\'{\i}sica,
Comisi\'on Nacional de Energ\'{\i}a At\'omica,\\
Av. del Libertador 8250, RA--1429 Buenos Aires, Argentina}
\affiliation{Departamento de F\'{\i}sica, Facultad de
Ciencias Exactas y Naturales,\\
Universidad de Buenos Aires, Ciudad Universitaria,
RA--1428 Buenos Aires, Argentina}
\affiliation{Consejo Nacional de Investigaciones
Cient\'{\i}ficas y T\'ecnicas,\\
Av. Rivadavia 1917, RA--1033 Buenos Aires, Argentina}

\date{\today}

\begin{abstract}
An extension of the nonlinear feedback (NLF) formalism to describe
regimes of hyper- and high-inflation in economy is proposed in the
present work. In the NLF model the consumer price index (CPI)
exhibits a finite time singularity of the type $1/(t_c -t)^{(1-
\beta)/\beta}$, with $\beta>0$, predicting a blow up of the
economy at a critical time $t_c$. However, this model fails in
determining $t_c$ in the case of weak hyperinflation regimes like,
e.g., that occurred in Israel. To overcome this trouble, the NLF
model is extended by introducing a parameter $\gamma$, which
multiplies all therms with past growth rate index (GRI). In this
novel approach the solution for CPI is also analytic being
proportional to the Gaussian hypergeometric function
$_2F_1(1/\beta,1/\beta,1+1/\beta;z)$, where $z$ is a function of
$\beta$, $\gamma$, and $t_c$. For $z \to 1$ this hypergeometric
function diverges leading to a finite time singularity, from which
a value of $t_c$ can be determined. This singularity is also
present in GRI. It is shown that the interplay between parameters
$\beta$ and $\gamma$ may produce phenomena of multiple equilibria.
An analysis of the severe hyperinflation occurred in Hungary
proves that the novel model is robust. When this model is used for
examining data of Israel a reasonable $t_c$ is got. High-inflation
regimes in Mexico and Iceland, which exhibit weaker inflations
than that of Israel, are also successfully described.
\end{abstract}


\maketitle

\section{Introduction}
\label{sec:introduction}

When inflation surpasses moderate levels it damages real economic
activities. For instance, it affects government's tax revenues
because they are effectively received a period later after the
declaration that fix them \cite{olivera67,tanzi77}. The perception
of relative prices changes becomes more difficult since it is not
easy to distinguish whether some price grows as a consequence of a
relative price change or it is part of general inflation (the
Lucas problem \cite{lucas72}). It produces inefficient changes of
relative prices \cite{king12} if the adjustment process is
different for different kinds of goods inducing misleading
allocation of resources \cite{fisher81}. Inflation also affects
currency in its property as medium of exchange, store of value,
and unit of account. The degree of perturbation is greater the
higher the inflation. If consumption goods become relatively more
expensive than leisure due to inflation, labor market may be
negatively perturbed by reduction of working hours supply
\cite{cooley89}. Unanticipated increase in prices reduces real
wages and expand employment \cite{taylor80}, although the positive
employment effect could be lowered or reversed by the effects of
falling investment. Furthermore, investment demand may be
especially affected because of the shorter planning time scope and
growing uncertainty. In general, when the inflation is higher then
the decision horizon is shorter. Moreover, with no alternative
allocation to money, savings decline and investment falls at
expenses of actual consumption, lowering the capital stock growth.
Therefore inflation is not a pure nominal problem,m but it is
linked to real economy in many non trivial ways \cite{heymann95}.
Consequently, governments try to prevent high inflation, and to
lower it when reach elevated levels. Parameters can change once
policy changes \cite{lucas76}. The relation between sources of
inflation and the evolution of parameters is complex and not
direct. These issues are also analyzed in text books on
econophysics \cite{stanley99,stauffer99,sornette03b}.

Models of hyperinflation are especially suitable to emphasize that
inflation implies bad ``states of nature'' in economy. Wars,
changes of social regimens, states bankruptcies are the
characteristics of such regimens. These factors together with the
influence of pure economical variables like expectations, money
demand, velocity of circulation and quantity of money, give an
increase of the consumer price index (CPI) larger than
exponential as can be observed in the investigated cases
\cite{cagan56,mizuno02,sornette03,szybisz08,szybisz09,szybisz10,%
szybisz15,szybisz16}. In turn, this behavior affects negatively
the social network causing unpleasant situations.

The model for hyperinflation available in the literature is based
on a nonlinear feedback (NLF) characterized by an exponent
$\beta>0$ of a power law. In such an approach the CPI exhibits a
finite time singularity of the form $1/(t_c-t)^{(1-\beta)/\beta}$,
allowing a determination of a critical time $t_c$ at which the
economy would crash. This model has been successfully applied to
many cases \cite{sornette03,szybisz08,szybisz09,szybisz10,%
szybisz15,szybisz16}. However, in the most recent paper
\cite{szybisz16}, it is shown that for the episode of weak
hyperinflation occurred in Israel it is impossible to determine a
value for $t_c$ within the NLF model because $\beta$ goes to zero.
In this limit one gets the linear feedback approach which does not
contain information on $t_c$ \cite{mizuno02}. This drawback was
attributed to a permanent but partially successful efforts for
stopping inflation in Israel.

In order to include in the NLF formalism information on efforts
for stabilization like those observed in the case of Israel, we
developed an extension of this model introducing a parameter
$\gamma$, which multiplies all the past growth rate index (GRI)
contributions changing the relative weight of the therm with the
power law. The interplay between $\beta$ and $\gamma$ leads to
multiple equilibria phenomena for episodes of high inflation. The
literature on this kind of behavior in models of economics is
large and diverse \cite{cooper99}, see also e.g. Refs.\
\cite{diamond82,bruno1987,obstfeld96,krugman99,morris00,nadal05}.
We use the well known multiple equilibria expression to indicate
the existence of multiple trajectories compatible with data which
lead to very different states of nature of the economy where the
final outcome is stable or not. In the extended NLF approach the
solution for CPI becomes proportional to the Gaussian
hypergeometric function $_2F_1(1/\beta,1/\beta,1+1/\beta;z)$,
where $z$ is a function of $\beta$, $\gamma$, and $t_c$. Since for
$z \to 1$ this hypergeometric function diverges, a finite time
singularity shows up allowing a determination of $t_c$. It is
important to notice that the Gaussian hypergeometric appears in a
variety of physical and mathematical problems. In quantum
mechanics, the solution of the Schr\"{o}dinger equation for some
potentials is expressed in terms of $_2F_1$ \cite{flugge71}.
Moreover, the eigenfunctions of the angular momentum operators are
sometimes written in terms of $_2F_1$ functions
\cite{abramowitz72}.

In the present work we also investigated the effects of the
parameter $\gamma$ on the non linear dynamic evolution of real
inflations. Firstly, the robustness of the novel model is tested
analyzing the severe hyperinflation occurred in Hungary after
World War II. Next, it is successfully applied for studying the
weaker hyperinflation developed in Israel and the high-inflation
episodes observed in Mexico and Iceland. In order to understand
better the evolution of prices in these countries, brief
descriptions of historical events are provided. In these cases
we found multiple equilibria.

The paper is organized in the following way. In Sec.\
\ref{sec:theory} the NLF model is outlined with some details
because it is the starting point for the extension proposed in
Sec.\ \ref{sec:general}. The limit $\gamma \to 1$ is discussed.
Section \ref{sec:results} is devoted to report and discuss the
results obtained by applying the novel approach. Finally, the
main conclusions are summarized in Sec.\ \ref{sec:summary}.

\section{Theoretical background}
\label{sec:theory}

Let us recall that the rate of inflation $i(t)$ is defined as
\begin{equation}
i(t) = \frac{P(t)-P(t-\Delta t)}{P(t-\Delta t)}
= \frac{P(t)}{P(t-\Delta t)} - 1  \:, \label{infla}
\end{equation}
where $P(t)$ is the CPI at time $t$ and $\Delta t$ is the period
of the measurements. In the academic financial literature, the
simplest and most robust way to account for inflation is to take
logarithm. Hence, the continuous rate of change in prices is
defined as
\begin{equation}
C(t) = \frac{\partial \ln{P(t)}}{\partial t} \:. \label{c_rate0}
\end{equation}
Usually the derivative of Eq.\ (\ref{c_rate0}) is expressed in a
discrete way as
\begin{eqnarray}
C(t+\frac{\Delta t}{2}) &=& \frac{\left[ \ln P(t+\Delta t)-
\ln P(t) \right]}{\Delta t} \nonumber\\
&=& \frac{1}{\Delta t}\,\ln \left[ \frac{P(t+\Delta t)}{P(t)}
\right] \:. \label{c_rate1}
\end{eqnarray}
The GRI over one period is defined as
\begin{eqnarray}
r(t+\frac{\Delta t}{2}) &\equiv& C(t+\frac{\Delta t}{2})\,\Delta t
= \ln \left[ \frac{P(t+\Delta t)}{P(t)} \right] \nonumber\\
&=& \ln[1 + i(t+\Delta t)] = p(t+\Delta t) - p(t) \:, \nonumber\\
\label{rate1}
\end{eqnarray}
where a widely utilized notation
\begin{equation}
p(t) = \ln P(t) \:, \label{plog}
\end{equation}
was introduced. It is straightforward to show that the accumulated
CPI is given by
\begin{equation}
P(t) = P(t_0)\,\exp{\left[\frac{1}{\Delta t}\int^t_{t_0} r(t') dt'
\right]} \;. \label{pt}
\end{equation}

\subsection{Adaptive inflationary expectation}
\label{sec:cagan}

In the fifties, Cagan has proposed \cite{cagan56} a model of
inflation based on the mechanism of ``adaptive inflationary
expectation'' with positive feedback between realized growth of
the market price $P(t)$ and the growth of people's averaged
expectation price $P^*(t)$. These two prices are thought to evolve
due to a positive feedback mechanism: an upward change of market
price $P(t)$ in a unit time $\Delta t$ induces a rise in the
people's expectation price $P^*(t)$, and such an anticipation
pushes on the market price. So, one may write
\begin{equation}
\frac{P(t+\Delta t)}{P(t)} = \frac{P^*(t)}{P(t)}
= \frac{P^*(t)}{P^*(t-\Delta t)} \:. \label{cag1}
\end{equation}
Actually $P^*(t)/P(t)$ indicates that the process induces a non
exact proportional response of adaptation due to the fact that the
expected inflation $P^*(t)$ expands the response to the price
level $P(t)$ in order to forecast and meet the inflation of the
next period. Now, one may introduce the expected GRI
\begin{equation}
r^*(t+\frac{\Delta t}{2}) \equiv  C^*(t+\frac{\Delta t}{2})\,
\Delta t = \ln \left[ \frac{P^*(t+\Delta t)}
{P^*(t)} \right] \:. \label{rate2}
\end{equation}

\subsection{The NLF Model}
\label{sec:feedback}

Sornette, Takayasu, and Zhou (STZ) \cite{sornette03} introduced a
nonlinear feedback (NLF) process in the formalism suggesting that
the people's expectation price $P^*(t)$ obeys
\begin{eqnarray}
\ln \left[ \frac{P^*(t+\Delta t)}{P^*(t)} \right]
&=& \ln \left[ \frac{P(t)}{P(t-\Delta t)} \right] \nonumber\\
&\times& \left( 1 + 2\,a_p \biggr\{
\ln \left[\frac{P(t)}{P(t-\Delta t)}\right]\biggr\}^\beta \right)
\:. \nonumber\\
\label{rate00}
\end{eqnarray}
Here $a_p$ is a positive dimensionless feedback's strength and
$\beta > 0$ is the exponent of the power law. This expression
together with Eq.\ (\ref{cag1}) leads to
\begin{equation}
r(t+\Delta t) = r(t-\Delta t) + 2\,a_p\,[r(t-\Delta t)]^{1+
\beta} \;. \label{rate42}
\end{equation}
Taking the continuous limit of this relation one obtains the
following equation for the time evolution of $r$
\begin{equation}
\frac{dr}{dt} = \frac{a_p}{\Delta t}\,[r(t)]^{1+\beta} \;.
\label{rate43}
\end{equation}
The solution for GRI follows a power law exhibiting a singularity
at finite-time $t_c$ \cite{sornette03,szybisz08,szybisz09}
\begin{equation}
r(t) = r_0\,\biggr[\frac{1}{1-\beta\,a_p\,r_0^\beta
\left(\frac{t-t_0}{\Delta t}\right)} \biggr]^{1/\beta}
= r_0\,\left(\frac{t_c-t_0}{t_c-t}\right)^{1/\beta} \;.
\label{r_time}
\end{equation}
The critical time $t_c$ being determined by the initial GRI $r(t=
t_0)=r_0$, the exponent $\beta$, and the strength parameter $a_p$
\begin{equation}
\frac{t_c - t_0}{\Delta t} = \frac{1}{\beta\,a_p\,r_0^\beta} \;.
\label{c_time}
\end{equation}
The CPI is obtained by integrating $r(t)$ according to Eq.\
(\ref{pt}). For $\beta \ne 1$, denoted as case (i), it becomes
\begin{eqnarray} 
&&\ln \left[\frac{P(t)}{P_0}\right] = p(t) - p_0 = \int^t_{t_0}
r(t') \frac{dt'}{\Delta t} \nonumber\\
&=& \frac{r_0^{1-\beta}}{(1-\beta)\,a_p}
\biggr\{ \biggr[\frac{1}{1-\beta\,a_p\,r_0^\beta
\left(\frac{t-t_0}{\Delta t}\right)}
\biggr]^{\frac{1-\beta}{\beta}} - 1 \biggr\} \;. \nonumber\\
\label{lptn}
\end{eqnarray}

(i.a) For $0<\beta<1$ one also gets a finite-time singularity in
the $\log$-CPI according to the power law
\begin{equation}
p(t) = p_0 + \frac{r_0\,\beta}{1-\beta}
\left(\frac{t_c-t_0}{\Delta t}\right)
\biggr[\left(\frac{t_c-t_0}{t_c-t}\right)^{\frac{1-\beta}{\beta}}
- 1 \biggr] \;. \label{price1}
\end{equation}
This expression has been used for the analysis of hyperinflation
episodes reported in previous papers \cite{sornette03,szybisz09,%
szybisz08,szybisz10,szybisz15}.

(i.b) For $\beta>1$ one gets a finite-time singularity for $r(t)$,
but the log-CPI evolves as
\begin{equation}
p(t) = p_0 + \frac{r_0\,\beta}{\beta-1}
\left(\frac{t_c-t_0}{\Delta t}\right)
\biggr[1 - \left(\frac{t_c-t}{t_c-t_0}\right)^{\frac{\beta-1}
{\beta}} \biggr] \label{price3}
\end{equation}
converging to a finite value when time approaches the critical
value $t_c$
\begin{equation}
p(t \to t_c) = p_0 + \frac{r_0\,\beta}{\beta-1}\,
\left(\frac{t_c-t_0}{\Delta t}\right) \;. \label{p_lim}
\end{equation}

(ii) For $\beta=1$ the integration in Eq.\ (\ref{pt}) yields
\begin{equation}
\ln \left[\frac{P(t)}{P_0}\right] = \int^t_{t_0} r(t')
\frac{dt'}{\Delta t} = \frac{1}{a_p}
\ln \biggr[ \frac{1}{1-a_p\,r_0\,\left(\frac{t-t_0}{\Delta t}
\right)} \biggr] \;. \label{logdiv}
\end{equation}
Upon introducing $t_c$ in this expression, the time dependence of
$p(t)$ exhibits a logarithmic divergence
\begin{equation}
p(t) = p_0 + r_0 \left(\frac{t_c-t_0}{\Delta t}\right)
\ln \left(\frac{t_c-t_0}{t_c-t}\right) \;. \label{price2}
\end{equation}
It was found that for severe episodes of hyperinflation one may
get $\beta \to 1$ \cite{szybisz10,szybisz15}.

In both regimes, (i.a) and (ii), the CPI exhibits a finite-time
singularity at the same critical value $t_c$ as GRI. Hence, these
solutions correspond to a genuine divergence of $\ln P(t)$.

It is important to notice that for $\beta=0$ one gets the linear
feedback (LF) model suggested previously by Mizuno, Takayasu, and
Takayasu (MTT) \cite{mizuno02}. In this limit one arrives at
\begin{equation}
\frac{dr}{dt} = \frac{a_p}{\Delta t}\,r(t)~~~~\to~~~~r(t) = r_0\,
\exp \biggr[a_p\,\biggr(\frac{t-t_0}{\Delta t} \biggr)\,\biggr]
\;. \label{r_mizu}
\end{equation}
Hence, the CPI grows as a function of $t$ following a
double-exponential law \cite{szybisz09,mizuno02}
\begin{equation}
\ln P(t) = p(t) = p_0 + \frac{r_0}{a_p}\,\biggr\{ \exp \biggr[
a_p\,\biggr(\frac{t-t_0}{\Delta t} \biggr)\,\biggr] - 1 \biggr\}
\;. \label{p_mizu}
\end{equation}
In the LF model no $t_c$ can be determined. Furthermore, by
setting now $a_p=0$ one gets
\begin{equation}
\frac{dr}{dt} = 0~~~~\to~~~~r(t) = r_0\, \;,
\label{r_cagan}
\end{equation}
and
\begin{equation}
\ln P(t) = p(t) = p_0 + r_0\,\biggr(\frac{t-t_0}{\Delta t} \biggr)
\;, \label{p_cagan}
\end{equation}
which is the Cagan's original proposal.

\section{Extension of the NLF model}
\label{sec:general}

Let us now present a way for introducing to some extent
information on saturation within the framework of the theory
outlined in Sec.\ \ref{sec:feedback}. A simple generalization of
the formalism would be to include a parameter $\gamma$ multiplying
$r(t)$ in the feedback term
\begin{equation}
\frac{dr}{dt} = \frac{a_p}{\Delta t}\,[\gamma\,r(t)]^{1+\beta} \;.
\label{rate43_g0}
\end{equation}
However, this attempt leads to a mere change of the parameter
$a_p$ by $a_p\,[\gamma]^{1+\beta}$
\begin{equation}
\frac{dr}{dt} = \frac{a_p\,[\gamma]^{1+\beta}}{\Delta t}\,
[r(t)]^{1+\beta} \;. \label{rate43_g1}
\end{equation}
Hence, this kind of approach should be done in a more elaborated
way. An adequate procedure is to multiply by $\gamma$ all the
terms corresponding to past GRI, i.e. $\ln \left[ \frac{P(t)}{P(t-
\Delta t)} \right]$, in Eq.\ (\ref{rate00})
\begin{eqnarray}
\ln \left[ \frac{P^*(t+\Delta t)}{P^*(t)} \right]
&=& \gamma \ln \left[ \frac{P(t)}{P(t-\Delta t)} \right]
\nonumber\\
&\times& \left( 1 + 2\,a_p \biggr\{ \gamma
\ln \left[\frac{P(t)}{P(t-\Delta t)}\right]\biggr\}^\beta \right)
\:. \nonumber\\
\label{rategg}
\end{eqnarray}
In this case the Eq.\ (\ref{rate42}) becomes
\begin{equation}
r(t+\Delta t) = \gamma\,r(t-\Delta t) + 2\,a_p\,[\gamma\,
r(t-\Delta t)]^{1+\beta} \;. \label{rate45}
\end{equation}
The new parameter $\gamma$ would account for the changes of
actions in the private, external and/or government sector.
Equation (\ref{rate45}) may be cast into the form
\begin{eqnarray}
r(t+\Delta t) &-& r(t-\Delta t) = (\gamma-1)\,r(t-\Delta t)
\nonumber\\
&&+ 2\,a_p\,\gamma^{1+\beta}\,[r(t-\Delta t)]^{1+\beta} \;,
\label{rate46}
\end{eqnarray}
which can be rewritten as
\begin{eqnarray}
r(t+2\Delta t) - r(t) = (\gamma-1)\,r(t)
+ 2\,a_p\,\gamma^{1+\beta}\,[r(t)]^{1+\beta} \;.
\nonumber\\
\label{rate47}
\end{eqnarray}
Upon dividing both sides of this relation by $2\Delta t$ and doing 
the transition from discrete to continues functions one gets a
differential equation of the Bernoulli type
\begin{eqnarray}
\frac{dr}{dt} &=& \dot{r}(t) = \frac{\gamma-1}{2\,\Delta t}\,r(t)
+ \frac{a_p}{\Delta t}\,\gamma^{1+\beta}\,[r(t)]^{1+\beta}
\nonumber\\
&=& s\,r(t) + q\,[r(t)]^{1+\beta} \;, \label{rate48}
\end{eqnarray}
where
\begin{equation}
s = \frac{\gamma-1}{2\,\Delta t}~~~~and~~~~q = \frac{a_p}
{\Delta t}\,\gamma^{1+\beta} \;, \label{def_sq}
\end{equation}
are introduced to simplify the notation at this stage. Equation
(\ref{rate48}) can be cast into the form
\begin{equation}
\frac{\dot{r}(t)}{[r(t)]^{1+\beta}} - \frac{s}{[r(t)]^\beta} = q
\;. \label{rate49}
\end{equation}
Introducing the change of variables
\begin{equation}
v(t) = \frac{1}{[r(t)]^\beta}~~~~and~~~~\dot{v}(t) =
- \beta\,\frac{\dot{r}(t)}{[r(t)]^{1+\beta}} \;, \label{rate50}
\end{equation}
one gets the linear differential equation
\begin{equation}
\dot{v}(t) + \beta\,s\,v(t) = -\beta\,q \;, \label{rate51}
\end{equation}
with the constrain
\begin{equation}
v(t=t_0) = v_0 = \frac{1}{[r(t=t_0)]^\beta} = \frac{1}{r_0^\beta}
\;. \label{rate52}
\end{equation}
The solution is
\begin{equation}
v(t) = -\frac{q}{s} + \biggr(\frac{1}{r_0^\beta}+\frac{q}{s}
\biggr)\,\exp[-\beta\,s\,(t-t_0)] \;, \label{rate53}
\end{equation}
which leads to
\begin{eqnarray}
\frac{1}{[r(t)]^\beta} &=& \frac{1}{r_0^\beta}
\biggr[\,\exp[-\beta\,s\,(t-t_0)] \nonumber\\
&& + \frac{q\,r_0^\beta}{s}\,\biggr\{\exp[-\beta\,s\,(t-t_0)]
- 1 \biggr\} \biggr] \;. \label{rate54}
\end{eqnarray}
Hence, the GRI in this extended NLF (denoted as ENLF) model
becomes
\begin{eqnarray}
r(t) &=& r_0 \biggr[\biggr( 1 + \frac{q\,r_0^\beta}{s} \biggr)\,
\exp[-\beta\,s\,(t-t_0)] - \frac{q\,r_0^\beta}{s}\,
\biggr]^{-1/\beta} \nonumber\\
&=& r_0\,\exp[s\,(t-t_0)] \nonumber\\
&&~~\times \biggr[\frac{s}{s + q\,r_0^\beta
( 1 - \exp[\beta\,s\,(t-t_0)]) } \biggr]^{1/\beta} \;.
\label{rate55}
\end{eqnarray}
In the limit $\gamma \to 1$ one gets
\begin{equation}
s \to 0~~~~and~~~~q \to \frac{a_p}{\Delta t} = 1/[\beta\,
r^\beta_0\,(t_c-t_0)] \;. \label{qlim}
\end{equation}
In turn, the CPI is given by
\begin{eqnarray} 
\ln \left[\frac{P(t)}{P_0}\right] &=& r_0 \int^t_{t_0}
\frac{dt'}{\Delta t} \, \exp[s\,(t'-t_0)] \nonumber\\
& \times & \biggr[\frac{s}{s + q\,r_0^\beta
( 1 - \exp[\beta\,s\,(t'-t_0)]) } \biggr]^{1/\beta} \;.
\nonumber\\
\label{lptng}
\end{eqnarray}
The general solution $\forall \beta$ provided by the Wolfram's
Mathematica on line integrator reads
\begin{eqnarray} 
\ln \left[\frac{P(t)}{P_0}\right] &=& \frac{r_0}{s \Delta t}\,
\biggr( \frac{s}{s + q r_0^\beta(1 - \exp[\beta s (t'-t_0)])}
\biggr)^{1/\beta} \nonumber\\
&& \times \exp[s (t'-t_0)] \, \biggr(1 - z \biggr)^{1/\beta}
\nonumber\\
&& \times \,_2F_1(1/\beta,1/\beta,1+1/\beta;z) \biggr|^t_{t_0} \;,
\label{lptng1}
\end{eqnarray}
where $_2F_1(1/\beta,1/\beta,1+1/\beta;z)$ is the Gaussian
hypergeometric function (see Appendix A) with
\begin{equation}
z = \frac{q\,r_0^\beta \exp[\beta s (t'-t_0)]}{s + q\,r_0^\beta}
\;. \label{defx}
\end{equation}
Furthermore, by introducing this expression for $z$ in Eq.\
(\ref{lptng1}), the latter equation for log-CPI can be cast into a
more compact form
\begin{eqnarray} 
\ln \left[\frac{P(t)}{P_0}\right] &=& \frac{r_0}{s\,\Delta t}\,
\biggr(\frac{s}{s + q\,r_0^\beta} \biggr)^{1/\beta}
\exp[s(t'-t_0)] \nonumber\\
&& \times \,_2F_1(1/\beta,1/\beta,1+1/\beta;z) \biggr|^t_{t_0}
\nonumber\\
&=& \frac{r_0}{s \Delta t}
\biggr( \frac{s}{s + q r_0^\beta} \biggr)^{1/\beta} \nonumber\\
& \times & \biggr[\,\exp[s(t-t_0)]
\,_2F_1(1/\beta,1/\beta,1+1/\beta;z) \nonumber\\
&& - \, _2F_1(1/\beta,1/\beta,1+1/\beta;z_0) \biggr] \;,
\label{lptng3}
\end{eqnarray}
with
\begin{equation}
z_0 = z(t=t_0) = \frac{q\,r_0^\beta}{s + q\,r_0^\beta} \;,
\label{defx0}
\end{equation}
and
\begin{equation}
z = z_0\,\exp[\beta s (t-t_0)] \;. \label{defxg}
\end{equation}
Notice that for $z \to 1$ the hypergeometric function
$_2F_1(1/\beta,1/\beta,1+1/\beta;z)$ diverges. Let us emphasize
that Eq.\ (\ref{lptng3}) embodies solutions or all sectors of
$\beta$. Although for $\beta=1$ a simple integration yields
\begin{eqnarray}
\ln \left[\frac{P(t)}{P_0}\right] &=& r_0 \, \int^t_{t_0}
\frac{dt'}{\Delta t} \, \exp[s\,(t'-t_0)] \nonumber\\ 
&& \times \biggr[\frac{s}{s + q\,r_0
( 1 - \exp[s\,(t'-t_0)]) } \biggr] \nonumber\\
&=& -\,\frac{\ln [s+q\,r_0\,(1-\exp[ s (t'-t_0)])]}{q\,\Delta t} 
\biggr|^t_{t_0} \nonumber\\
&=& \frac{1}{q\,\Delta t}
\ln \biggr[ \frac{s}{s+q\,r_0\,(1-\exp[ s (t-t_0)])} \biggr] \;.
\nonumber\\
\label{lptln1}
\end{eqnarray}
The latter result can be also obtained by introducing into Eq.\
(\ref{lptng3}) the expression of the hypergeometric functions
corresponding to $\beta=1$
\begin{equation}
_2F_1(1,1,2;z) = - \frac{1}{z} \ln (1 - z) \;. \label{2F1b1}
\end{equation}
This procedure yields
\begin{eqnarray}
\ln \left[\frac{P(t)}{P_0}\right] &=& \frac{r_0}{\Delta t}
\biggr( \frac{1}{s + q r_0} \biggr) \frac{1}{z_0} \nonumber\\
& \times & \biggr[- \ln (1-z) + \ln (1-z_0) \biggr]
\;, \label{lptngb1}
\end{eqnarray}
leading to Eq.\ (\ref{lptln1}).

Equations (\ref{lptng3}) and (\ref{lptln1}) have two domains of 
solutions, one for $\gamma > 1$ (implying $s > 0$) and the other
for $\gamma < 1$ ($s < 0$). In the Appendix B we show that in the
limiting case $\gamma \to 1$ (i.e., $s \to 0$) from above, the
generalized expressions for GRI and CPI provided by the ENLF model
reduce to the forms reported previously in Refs.\
\cite{szybisz09,sornette03} and summarized in Sec.\
\ref{sec:theory}.

\subsection{Critical time $t_c$ in the ENLF model}
\label{sec:critical_t}

Let us now rewrite Eqs.\ (\ref{rate55}), (\ref{lptng3}), and
(\ref{lptln1}) in terms of $\gamma$ and determine the critical
time $t_c$ within this novel formalism. In so doing, after
inserting the definitions of $s$ and $q$ in Eq.\ (\ref{rate55})
for GRI, we obtain
\begin{eqnarray}
&&r(t) = r_0 \nonumber\\
&& \times \biggr[\frac{\gamma-1}{(\gamma -1 + 2 a_p
\gamma^{1+\beta} r_0^\beta) \exp(-\delta) - 2 a_p \gamma^{1+\beta}
r_0^\beta } \biggr]^{1/\beta} \;, \nonumber\\
\label{rate56}
\end{eqnarray}
with 
\begin{equation}
\delta = \beta\,\frac{(\gamma-1)}{2}\,
\biggr(\frac{t-t_0}{\Delta t} \biggr) \;. \label{deltaa}
\end{equation}
The CPI becomes $\forall \beta$
\begin{eqnarray} 
\ln \left[\frac{P(t)}{P_0}\right] &=& \frac{2\,r_0}{\gamma -1}
\biggr( \frac{\gamma -1}{\gamma -1 + 2\,a_p\,\gamma^{1+\beta}\,
r_0^\beta} \biggr)^{1/\beta} \nonumber\\
&& \times \biggr[\,\exp(\delta/\beta)
\,\,_2F_1(1/\beta,1/\beta,1+1/\beta;z) \nonumber\\
&&~~~ - \, _2F_1(1/\beta,1/\beta,1+1/\beta;z_0) \biggr] \;,
\label{lptng4}
\end{eqnarray}
with
\begin{equation}
z = \frac{2\,a_p\,\gamma^{1+\beta}\,r_0^\beta \, \exp(\delta)}
{\gamma -1 + 2\,a_p\,\gamma^{1+\beta}\,r_0^\beta} \;, 
\label{defxa}
\end{equation}
keeping in mind that $z_0 = z(t=t_0)$. For $\beta = 1$, the CPI
reduces to
\begin{eqnarray}
\ln \left[\frac{P(t)}{P_0}\right] &=& \frac{1}{a_p\,\gamma^2}
\nonumber\\
&\times& \ln \biggr[ \frac{\gamma-1}{\gamma-1+2\,a_p\,\gamma^2\,
r_0 - 2\,a_p\,\gamma^2\,r_0\,\exp(\delta)} \biggr] \;. \nonumber\\
\label{lptln2}
\end{eqnarray}

The finite time singularity occurs at the same value of critical
$\delta_c$ for GRI and for both sectors (\ref{lptng4}) [with
(\ref{defxa})] and (\ref{lptln2}) of solutions for log-CPI,
satisfying 
\begin{equation}
\exp(\delta_c) = 1 + \frac{\gamma -1}{2\,a_p\,\gamma^{1+\beta}\,
r_0^\beta} \;. \label{singul_0}
\end{equation}
In turn, the critical time $t_c$ can be determined by equating
this relation with $\delta_c=\delta(t=t_c)$ given by Eq.\
(\ref{deltaa})
\begin{eqnarray}
\delta_c = \beta \frac{(\gamma-1)}{2}\biggr(\frac{t_c-t_0}
{\Delta t} \biggr) = \ln \biggr[ 1 + \frac{\gamma -1}{2 a_p
\gamma^{1+\beta} r_0^\beta} \biggr] \;. \label{singul_1}
\end{eqnarray}
The result is
\begin{equation}
\frac{t_c-t_0}{\Delta t} = \frac{2}{\beta\,(\gamma-1)}\,\ln
\biggr[ 1 + \frac{\gamma -1}{2\,a_p\,\gamma^{1+\beta}\,r_0^\beta}
\biggr] \;, \label{tc_g}
\end{equation}
providing the new expression for $t_c$.

Let us now show that in the limit $\gamma \to 1$ one retrieves for 
$t_c$ the expression given in Sect. \ref{sec:feedback}. Expanding
the logarithm in powers of $(\gamma -1)/(2\,a_p\,\gamma^{1+\beta}
\,r_0^\beta)$ one gets
\begin{eqnarray}
\frac{t_c-t_0}{\Delta t} &=& \biggr[ \frac{2}{\beta\,(\gamma-1)} 
\biggr] \biggr[ \frac{\gamma -1}{2\,a_p\,\gamma^{1+\beta}\,
r_0^\beta} \nonumber\\
&&-\,\frac{1}{2} \biggr( \frac{\gamma -1}{2\,a_p\,\gamma^{1+
\beta}\,r_0^\beta} \biggr)^2 + {\cal{O}}(h-o) \biggr] \nonumber\\
&\simeq & \frac{1}{\beta\,a_p\,\gamma^{1+\beta}\,r_0^\beta}
\biggr( 1 - \frac{\gamma -1}{4\,a_p\,\gamma^{1+\beta}\, r_0^\beta}
\biggr) \;. \label{tc_gl}
\end{eqnarray}
Then, for $\gamma \to 1$ one arrives at
\begin{equation}
\frac{t_c-t_0}{\Delta t} =  \frac{1}{\beta\,a_p\,r_0^\beta} \;,
\label{tc_gl1}
\end{equation}
recovering the relation of Eq.\ (\ref{c_time}).

\subsection{Observables as a function of $t_c$ in the ENLF model}
\label{sec:observables}

Upon introducing Eq.\ (\ref{singul_0}) into Eq.\ (\ref{rate56})
the GRI becomes  
\begin{eqnarray}
&&r(t) = r_0 \biggr[ \frac{\gamma-1}
{2\,a_p\,\gamma^{1+\beta}\,r_0^\beta } \biggr]^{1/\beta}
\nonumber\\
&&~~ \times \biggr[\biggr(1 + \frac{\gamma -1}{2 a_p
\gamma^{1+\beta} r_0^\beta } \biggr) \exp(-\delta) - 1
\biggr]^{-1/\beta} \nonumber\\
&=& r_0 \biggr[ \frac{\exp(\delta_c)-1}{\exp(\delta_c-\delta)-1}
\biggr]^{1/\beta}
= r_0 \biggr[ \frac{1-z_0}{\exp(-\delta)-z_0}\biggr]^{1/\beta}
\;, \nonumber\\ \label{rate57}
\end{eqnarray}
with
\begin{equation}
z_0 = \exp(-\delta_c)
= \exp \biggr[-\beta\,\frac{(\gamma-1)}{2}
\biggr(\frac{t_c-t_0}{\Delta t}\biggr) \biggr] \;. \label{defxb0}
\end{equation}

\begin{table*}
\caption{\label{tab:table3}Parameters obtained from the analysis
of the hyperinflation in Hungary after World War II.}
\begin{ruledtabular}

\begin{tabular}{lccccclc}
Period & \multicolumn{5}{c}{Parameters} & Model & $\chi$ \\
\cline{2-6}
       & $t_c$ & $r_0$ & $\beta$ & $\gamma$ & $p_0$ & \\
\hline
 \\
1945:04:30-1946:07:15 & 1946:09:03        & 0.150
& 0.500           &               & 3.82 & NLF$^*$ & 1.168 \\
                      & 1946:07:30$\pm$01 & 0.199$\pm$0.003
& 0.700$\pm$0.004 &                   & 4.07 & NLF & 0.759 \\
                      & 1946:07:28$\pm$04 & 0.216$\pm$0.017
& 0.733$\pm$0.044 &              & 3.66$\pm$0.17 & NLF & 0.704 \\
                      & 1946:07:30$\pm$01 & 0.199$\pm$0.003
& 0.700$\pm$0.004 & 1.0001$\pm$0.0058 & 4.07 & ENLF & 0.761 \\

1945:04:30-1946:07:31 & 1946:09:02$\pm$05 & 0.158$\pm$0.011
& 0.514$\pm$0.026 &                   & 4.07 & NLF  & 1.123 \\
                      & 1946:09:01$\pm$03 & 0.160$\pm$0.008
& 0.518$\pm$0.014 &              & 3.44$\pm$0.27 & NLF  & 1.123 \\
                      & 1946:09:02$\pm$01 & 0.158$\pm$0.001
& 0.514$\pm$0.001 & 1.0003$\pm$0.0179 & 4.07 & ENLF & 1.123 \\

\end{tabular}
$^*$ The values of the parameters listed in this line were
calculated using those reported by STZ \cite{sornette03}, see
text.
\end{ruledtabular}
\end{table*}

Furthermore, the CPI can be cast into the form
\begin{eqnarray} 
p(t) &=& p_0 + \frac{2\,r_0}{\gamma -1}\,
\biggr( 1 - z_0 \biggr)^{1/\beta} \nonumber\\
&&~~ \times \biggr[\,\exp(\delta/\beta)
\,_2F_1(1/\beta,1/\beta,1+1/\beta;z_0\,\exp(\delta)) \nonumber\\
&&~~~~ - \, _2F_1(1/\beta,1/\beta,1+1/\beta;z_0) \biggr]
\nonumber\\
&=& p_0 + \frac{2\,r_0}{\gamma -1}\,
\biggr( 1 - z_0 \biggr)^{1/\beta} \nonumber\\
&&~~ \times \biggr[\,\exp(\delta/\beta)
\,_2F_1(1/\beta,1/\beta,1+1/\beta;z) \nonumber\\
&&~~~~ - \, _2F_1(1/\beta,1/\beta,1+1/\beta;z_0) \biggr] \;,
\label{lptng5}
\end{eqnarray}
where $z$ is
\begin{equation}
z = z_0\,\exp(\delta)
= \exp \biggr[-\beta\,\frac{(\gamma-1)}{2}\,\biggr(
\frac{t_c-t}{\Delta t} \biggr) \biggr] \;. \label{defxb}
\end{equation}
For $t < t_c$ one gets $z < 1$, while at $t = t_c$ the value of
$z$ becomes unity and at this point the hypergeometric function
diverges. For $\beta = 1$, instead of Eq.\ (\ref{lptng5}) one can
write
\begin{eqnarray}
p(t) &=& p_0 + \frac{1}{a_p\,\gamma^2}
\biggr\{ \ln \biggr[ \frac{\gamma-1}{2\,a_p\,\gamma^2\,r_0}
\biggr] \nonumber\\
&& - \ln \biggr[ \biggr( 1 + \frac{\gamma-1}{2\,a_p\,\gamma^2\,
r_0} \biggr) - \exp(\delta) \biggr] \biggr\}\nonumber\\
&=& p_0 + \frac{2\,r_0}{\gamma-1} [\exp(\delta_c)-1]
\ln \biggr[ \frac{\exp(\delta_c)-1}{\exp(\delta_c) - \exp(\delta)} 
\biggr] \nonumber\\
&=& p_0 + \frac{2\,r_0}{\gamma-1} \biggr[ \frac{1 - z_0}{z_0}
\biggr] \ln \biggr[ \frac{1-z_0}{1-z} \biggr] \;. \label{lptln3}
\end{eqnarray}

\begin{figure*}
\includegraphics[width=8.5cm, height=6.5cm]{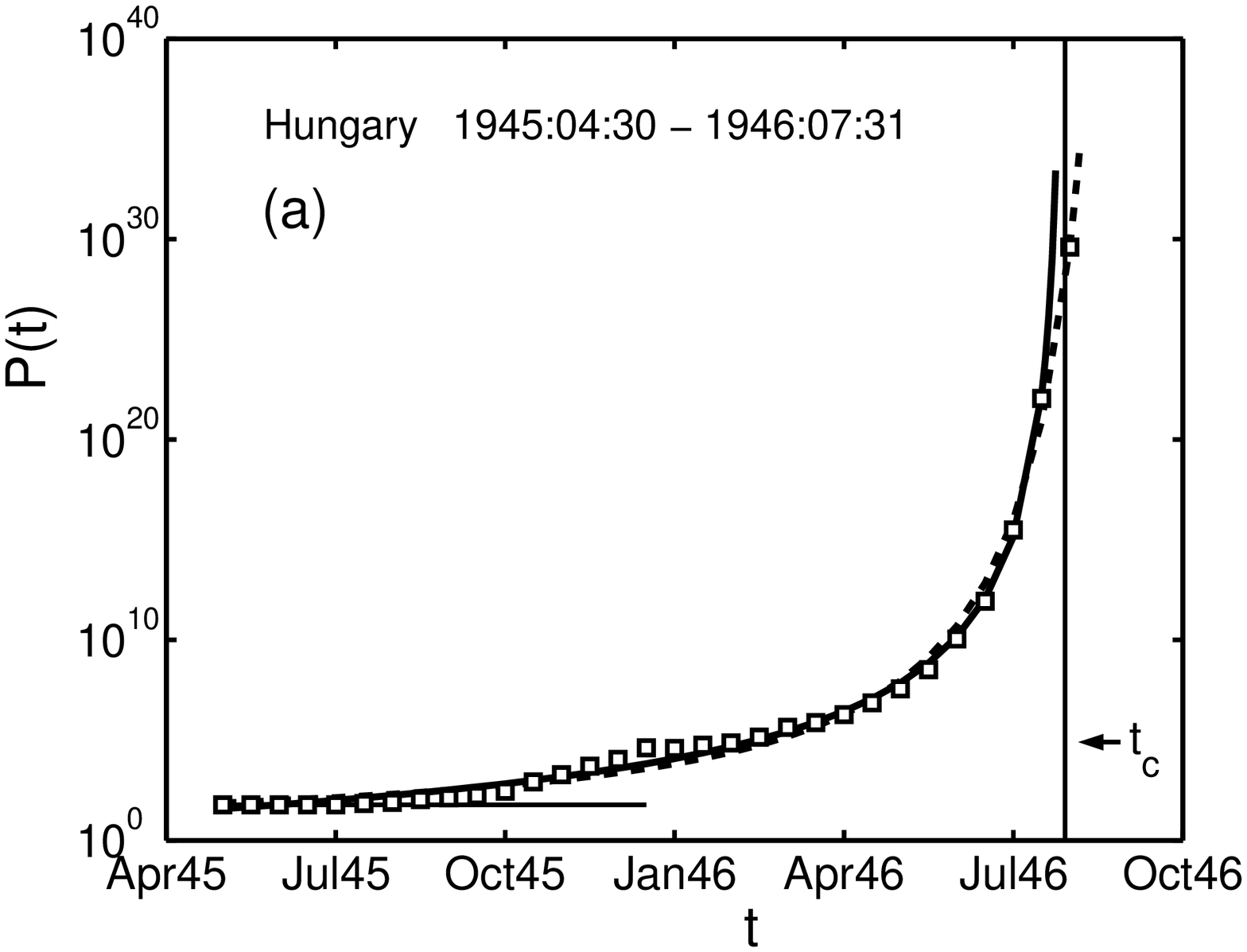}
\includegraphics[width=8.5cm, height=6.5cm]{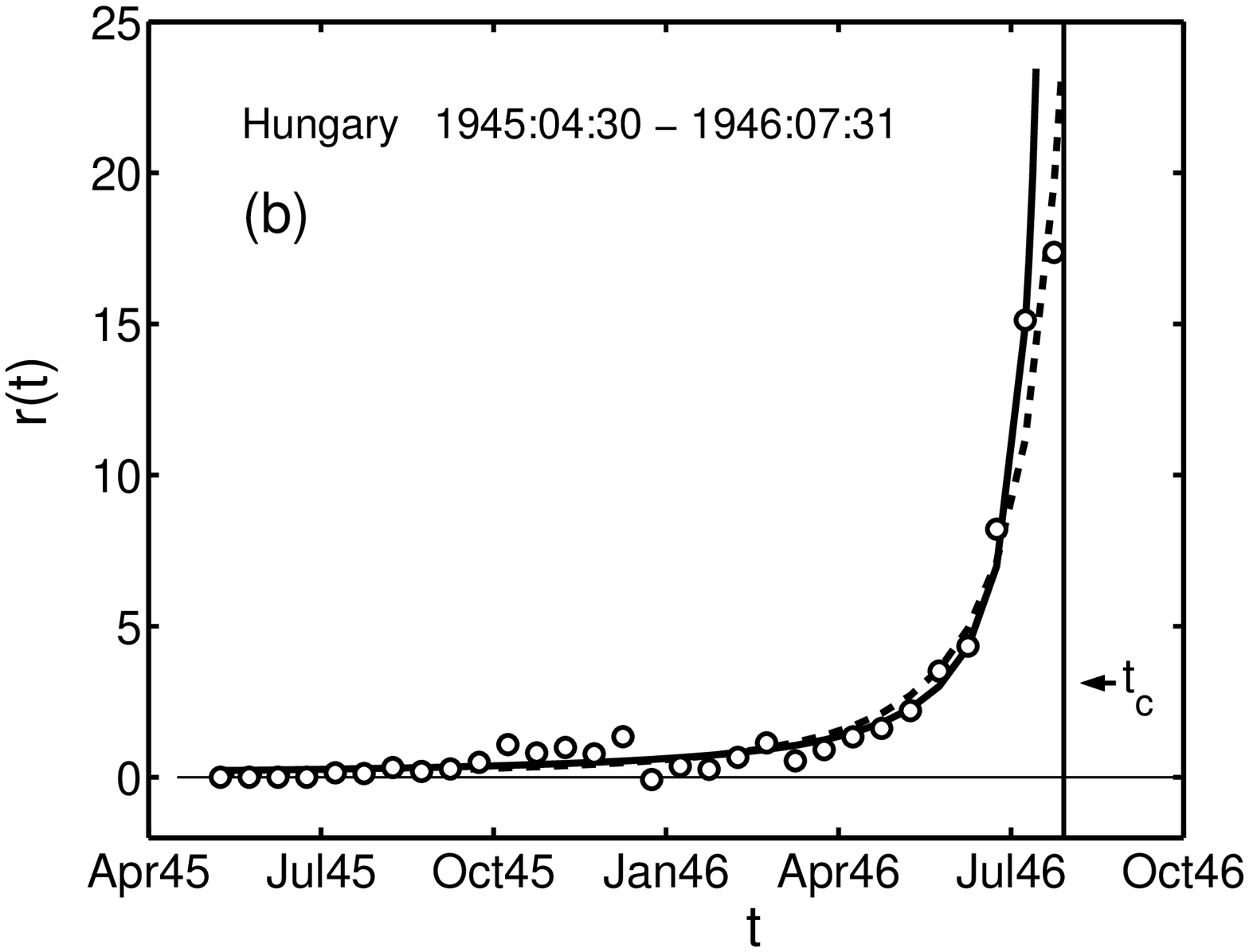}
\caption{\label{fig:Hungary}(a) Squares are data of biweekly CPI
in Hungary from 1945:04:30 to 1946:07:31 normalized to $P(t_0 =
1945:04:30)=58.35$. The horizontal continuous line indicates the
initial stable regime. (b) Circles are data of GRI for the same
period as in (a). Dashed curves in both panels indicate fits of
all these data to Eqs.\ (\ref{price1}) and (\ref{r_time}). Solid
curves stand for fits of data excluding the value at 1946:07:31.
The vertical solid lines are the predicted $t_c$ in the latter
case.}
\end{figure*}

\section{Analysis of hyper- and high-inflation episodes applying
the ENLF model}
\label{sec:results}

In a first step, we shall show that the ENLF formalism is robust
for episodes of severe hyperinflation, leading to values of
$\gamma$ very close to unity. Next, cases of weaker inflations
will be treated.

\subsection{Catastrophic hyperinflation in Hungary}
\label{sec:checking}

Let us begin the application of the novel approach by tackling an
emblematic case. An important test for any formalism developed for
describing regimes of high inflation is to verify whether it is
able to account for the evolution of prices occurred in Hungary
right after World War II \cite{mizuno02,sornette03}. For this
severe hyperinflation there is available a data series collected
in Table A.1 of Ref.\ \cite{paal00} (see also \cite{grossman00})
constructed on a basis of a biweekly frequency. The word
``biweekly'' is used to designate data available on the 15th and
on the last day of each month. Figure \ref{fig:Hungary} shows the
CPI and GRI during the period from April 30, 1945 to July 31, 1946
(i.e., from 1945:04:30 to 1946:07:31, in the case of Hungary the
notation Year:Month:Day will be used). Notice that the CPI values
are normalized to $P_0=P(t_0=1945:04:30)=58.35$. Several fits of
these data were performed.

An inspection to Fig.\ \ref{fig:Hungary}(b) suggests that the
value at 1946:07:31 could be already affected by the stabilization
policy adopted by the Hungarian government \cite{paal00}.
Therefore, in a first step we performed a fit of CPI data up to
1946:07:15 with Eq.\ (\ref{price1}) setting $p_0=\ln P_0=4.066$.
The numerical task was accomplished by using a routine of the book
by Bevington \cite{bevington69} cited as the first reference in
Chaps. 15.4 and 15.5 of the more recent {\it Numerical Recipes}
\cite{press96}. In practice, the applied procedure yields the
uncertainty in each parameter directly from the minimization
algorithm. The parameters yielded by these fits are listed in
Table \ref{tab:table3} together with the root-mean-square
(r.m.s.) residue of the fit, $\chi$. A simple inspection indicates
a crash at the beginning of August 1946. Values of GRI were
evaluated with Eq.\ (\ref{r_time}). The good quality of the fits
is shown by solid curves in panels (a) and (b) of Fig.\
\ref{fig:Hungary}.

Table 1 in Ref.\ \cite{sornette03} indicates that STZ have also
analyzed data of Hungary from 1945:04:30 to 1946:07:15. In order
to facilitate a quantitative comparison with the study reported by
STZ, we evaluated the parameters $r_0$, $\beta$, and $p_0$
utilized in the present work by inserting the values of $t_c$,
$\alpha$, $A$, and $B$ listed in Table 1 of Ref.\
\cite{sornette03} into the following relations
\begin{eqnarray}
&& r_0/\Delta t = \alpha\,B/(t_c-t_0)^{1+\alpha} \;, \label{r0}
 \\
&& \beta = 1/(1+\alpha) \;, \label{bet}
 \\
&& p_0 = A + B/(t_c-t_0)^\alpha \;. \label{p0}
\end{eqnarray}
These results are included in our Table\ \ref{tab:table3}. A
glance at this table indicates a shift between the present
results and that of STZ.

\begin{figure*}
\includegraphics[width=8cm, height=6cm]{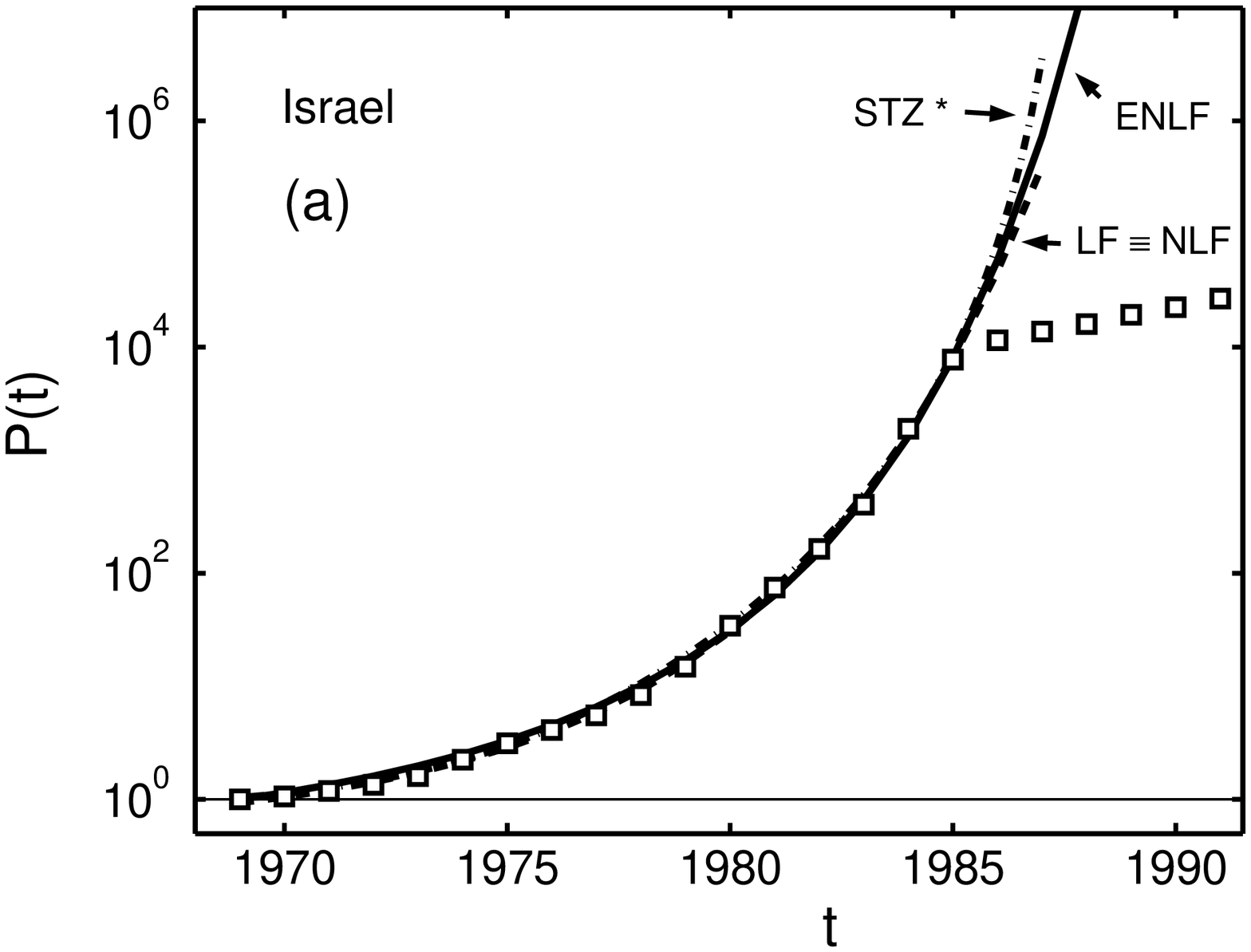}
\includegraphics[width=8cm, height=6cm]{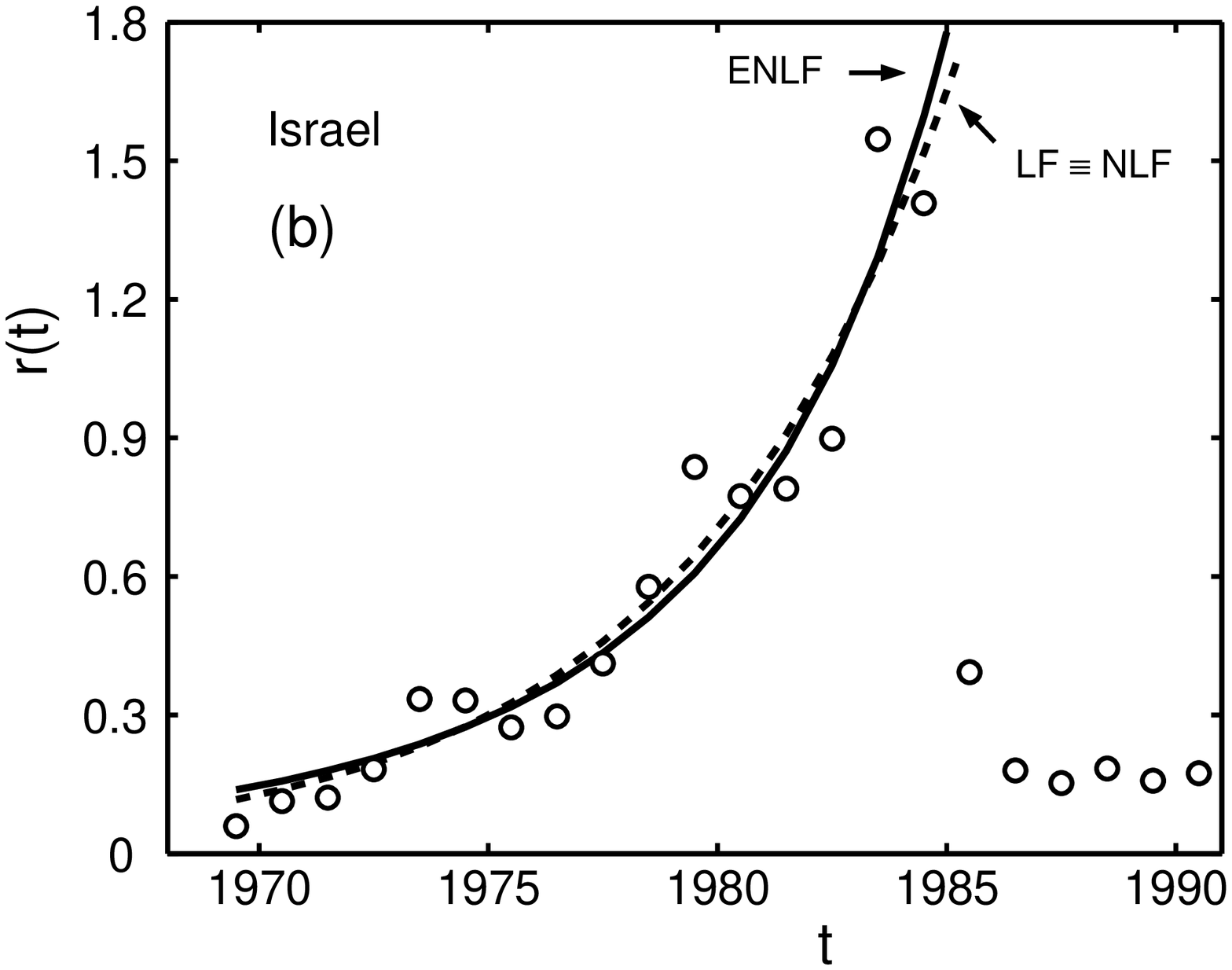}
\caption{\label{fig:Israel_h}(a) Open squares are yearly CPI in
Israel since 1969 to 1991 normalized to P($t_0$=1969)=1. The
dashed curve indicates the fit of $\ln P(t)$ since 1969 to 1985 to
Eq.\ (\ref{p_mizu}) corresponding to the LF model, while the
dot-dashed curve stands for the fit of $P(t)$ to Eq.\
(\ref{price1}) as reported by STZ (see text). The solid curve
stands for the fit of $\ln P(t)$ to Eq.\ (\ref{lptng5}) provided
by the present ENLF($t_c$) model. (b) Open circles are GRI for the
same period as in (a). The solid and dashed curves are evaluations
performed with Eqs.\ (\ref{rate57}) of the ENLF($t_c$) and
(\ref{r_mizu}) of the LF model, respectively.}
\end{figure*}

For the sake of completeness, we also analyzed CPI data including
the value at 1946:07:31. This fit yielded the parameters quoted in
Table \ref{tab:table3} and the dashed curves in panels (a) and (b)
of Fig.\ \ref{fig:Hungary} show the adjustment. As expected, the
new fit suggests a later date for the crash of the economy than
the prediction provided by the shorter series, now the blow up
would occur at the beginning of September 1946. Surprisingly, one
may observe in Table \ref{tab:table3} that the results of STZ,
including the $\chi$, are almost the same as that obtained in the
present work with the series ending at 1946:07:31. For the sake of
completeness, fits leaving $p_0$ free were also performed. Looking
at Table \ref{tab:table3} one may realize that no significant
changes are obtained.

As planned the study of the Hungarian hyperinflation is finished
by applying the ENLF formalism derived in Sec.\ \ref{sec:general}
of the present paper. In so doing, both the shorter and larger
series of CPI data were fitted to Eq.\ (\ref{lptng5}). The
obtained parameters are also quoted in Table \ref{tab:table3},
where one can observe that all the ``old'' parameters remain
unchanged. Predictions for GRI were computed using Eq.\
(\ref{rate57}). The match to Eqs.\ (\ref{rate57}) and
(\ref{lptng5}) and those corresponding to Eqs.\ (\ref{r_time}) and
(\ref{price1}) are indistinguishable on the scales of Fig.\
\ref{fig:Hungary}. One can realize that in both cases $\gamma$
stays equal to unity within the uncertainty. For the complete
series up to 1946:07:31 the uncertainty is slightly larger, this
feature could be attributed to the beginning of the stabilization
process. All these results support the robustness of the extended
model.

\begin{table*}
\caption{\label{tab:table4} Parameters obtained from the analysis
of episodes of high-inflation.}
\begin{ruledtabular}
\begin{tabular}{llccccccll}
Country & Period & \multicolumn{6}{c}{Parameters} & Model &
$\chi$ \\
\cline{3-8}
&& $t_c$ & $a_p$ & $r_0$ & $\beta$ & $\gamma$ & $z_0$ && \\
\hline
Iceland   & 1960-1983$^a$ & & 0.097$\pm$0.032 & 0.063$\pm$0.026
&   & & & LF & 0.0817$^b$ \\
          &               & 2049$\pm$122 & 0.120 & 0.068$\pm$0.027
& 0.136$\pm$0.225 & & & NLF & 0.0828$^b$ \\
          &               & 2141$\pm$264 & 0.107 & 0.065$\pm$0.026
& 0.061$\pm$0.096 & & & NLF & 0.0818$^c$ \\
          &           &     2291     & 0.076 & 0.064$\pm$0.051
& 0.037$\pm$1.664 & 1.044$\pm$0.073 & 0.765$\pm$0.391
& ENLF(z$_0$) & 0.0817$^b$ \\
          &             & 2049$\pm$18 & 0.092 & 0.067$\pm$0.014
& 0.151$\pm$0.033 & 1.042$\pm$0.066 & 0.774 & ENLF(t$_c$)
& 0.0825$^b$ \\
          &             & 2049.7$\pm$5.8 & 0.091 & 0.067$\pm$0.006
& 0.151$\pm$0.011 & 1.042$\pm$0.022 & 0.750 & ENLF(t$_c$)
& 0.0825$^c$ \\

Israel    & 1969-1985 & 1988.06 & & 0.077 & 0.149  & &
& STZ$^d$ & 0.085 \\
          &           & & 0.176$\pm$0.035 & 0.101$\pm$0.035
&   & & & LF & 0.0876$^b$ \\
          &           & 2061$\pm$72 & 0.184 & 0.109$\pm$0.035
& 0.069$\pm$0.061 & & & NLF & 0.0947$^b$ \\
          &           & 2527$\pm$456 & 0.177 & 0.102$\pm$0.035
& 0.010$\pm$0.009 & & & NLF & 0.0883$^c$ \\
          &           &     2170     & 0.130 & 0.104$\pm$0.016
& 0.022$\pm$0.117 & 1.076$\pm$0.008 & 0.778$\pm$0.022
& ENLF(z$_0$) & 0.0890$^b$ \\
          &           & 2015.5$\pm$4.8 & 0.146 & 0.116$\pm$0.018
& 0.172$\pm$0.021 & 1.069$\pm$0.068 & 0.758 & ENLF(t$_c$)
& 0.1046$^b$ \\
          &           & 2015.6$\pm$0.6 & 0.144 & 0.116$\pm$0.003
& 0.172$\pm$0.003 & 1.072$\pm$0.008 & 0.750 & ENLF(t$_c$)
& 0.1044$^c$ \\
          & 1969-1984 & & 0.178$\pm$0.045 & 0.100$\pm$0.040
&   & & & LF & 0.0885$^b$ \\
          &           & 2048$\pm$79  & 0.189 & 0.107$\pm$0.041
& 0.081$\pm$0.093 & & & NLF & 0.0942$^b$ \\
          &           & 2430$\pm$489 & 0.179 & 0.101$\pm$0.041
& 0.012$\pm$0.014 & & & NLF & 0.0892$^c$ \\
          &           & 2273         & 0.130 & 0.102$\pm$0.019
& 0.022$\pm$0.161 & 1.077$\pm$0.011 & 0.775$\pm$0.029
& ENLF(z$_0$) & 0.0895$^b$ \\
          &           & 2014.8$\pm$6.2 & 0.149 & 0.112$\pm$0.021
& 0.170$\pm$0.027 & 1.071$\pm$0.085 & 0.758 & ENLF(t$_c$)
& 0.0993$^b$ \\
          &           & 2014.9$\pm$0.7 & 0.147 & 0.112$\pm$0.003
& 0.170$\pm$0.003 & 1.074$\pm$0.010 & 0.750 & ENLF(t$_c$)
& 0.0992$^c$ \\

Mexico    & 1960-1988$^a$ & & 0.152$\pm$0.028 & 0.014$\pm$0.008
&   & & & LF & 0.0808$^b$ \\
          &               & 2132$\pm$99 & 0.162 & 0.016$\pm$0.009
& 0.043$\pm$0.028 & & & NLF & 0.0890$^b$ \\
          &               & 3175$\pm$731 & 0.153 & 0.015$\pm$0.008
& 0.006$\pm$0.004 & & & NLF & 0.0817$^c$ \\
          &           & 2069         & 0.127 & 0.018$\pm$0.008
& 0.082$\pm$0.149 & 1.061$\pm$0.014 & 0.763$\pm$0.052
& ENLF(z$_0$) & 0.0941$^b$ \\
          &             & 2033.6$\pm$1.3 & 0.139 & 0.020$\pm$0.002
& 0.134$\pm$0.005 & 1.058$\pm$0.014 & 0.751 & ENLF(t$_c$)
& 0.1045$^b$ \\
          &             & 2033.6$\pm$0.3 & 0.139 & 0.020$\pm$0.001
& 0.133$\pm$0.001 & 1.058$\pm$0.002 & 0.750 & ENLF(t$_c$)
& 0.1043$^c$ \\

\end{tabular}
$^a$ Data from Ref.\ \cite{FRED}.
$^b$ $\Delta \chi = \chi_i - \chi_{i+1} < 10^{-1}\%$.
$^c$ $\Delta \chi < 10^{-3}\%$.
$^d$ The values of the parameters listed in this line were
calculated using those reported by STZ \cite{sornette03} (see
text), in addition, $p_0$=1.04.
\end{ruledtabular}
\end{table*}

\subsection{Weak hyperinflation in Israel}
\label{sec:weak}

Let us now examine the case of Israel, which have been already
studied in a previous work \cite{szybisz16}. Figure\
\ref{fig:Israel_h}(a) shows the yearly data for the CPI in
Israel computed using data taken from a Table of the International
Monetary Fund (IMF) \cite{imf}. The upward curvature of the
logarithm of CPI as a function of time indicates that from 1969 to
1985 the hyperinflation exhibits a faster than exponential growth.
The inflation got triple-digit rates of about $400\%$ at their
peak in the mid-1980's. Leiderman and Liviatan \cite{leiderman03}
attributed this response to the implicit preference for short-term
considerations of avoiding unemployment over long-term monetary
stability. In 1985 a new strategy was applied to stop the
hyperinflation.

The results for the parameters and the $\chi$ obtained in the
previous work \cite{szybisz16} are reproduced in Table
\ref{tab:table4}. Two time series of CPI were examined: one with
data from 1969 until 1985 (this series has been studied in Refs.\
\cite{mizuno02} and \cite{sornette03}) and the other excluding the
value of 1985 when the final stabilization started. From that
analysis it was concluded that when applying the NLF model the
fits indicate a strong correlation between $\beta$ and $t_c$. In
Table \ref{tab:table4} two sets of parameters are provided, one
corresponds to stopping minimization when the change of $\chi$
from the ``$i$'' to the ``$i+1$''-iteration becomes less than
$\Delta \chi < 10^{-1} \%$ and the other for $\Delta \chi <
10^{-3} \%$. Let us mention that for that study $p_0$ was set
equal to zero. The fitting procedure showed that $\beta$ decreases
approaching zero while $t_c$ increases, this occurs in such a way
that the product $\beta \times (t_c-t_0)/\Delta t$ converges to a
constant yielding a well defined value of the parameter $a_p$
given by [see Eq.\ (\ref{c_time})]
\begin{equation}
a_p(\rm NLF) = \frac{\Delta t}{\beta\,r_0^\beta\,(t_c-t_0)} \;,
\label{a_p_lim}
\end{equation}
which is also quoted in Table \ref{tab:table4}. Since for $\beta
\to 0$ the NLF model converges towards the LF model of MTT
\begin{equation}
a_p(\rm MTT) = \lim_{\beta \to 0} \biggr[ \frac{\Delta t}
{\beta\,r_0^\beta\,(t_c-t_0)} \biggr]
= \frac{\Delta t}{\beta\,(t_c-t_0)}
\;, \label{a_p_ass}
\end{equation}
that study was completed by fitting the CPI data directly with
Eq.\ (\ref{p_mizu}). The obtained parameters are also quoted in
Table \ref{tab:table4}. Figure \ref{fig:Israel_h}(a) shows the
good quality of the fit for the larger series. It is also
worthwhile to mention that the present value for the parameter
utilized in the original LF model, i.e. $B_{\rm MTT}=1+2a_p=
1.352$, is in good agreement with the result $1.4$ quoted by MTT
in Table 1 of Ref.\ \cite{mizuno02}. For the sake of completeness,
we plotted in Fig.\ \ref{fig:Israel_h}(b) the measured data of GRI
together with the theoretical values provided by Eqs.\
(\ref{r_time}) and (\ref{r_mizu}), which are indistinguishable on
the scale of the drawing.

Although the LF model provides a good fit, it does not predict any
$t_c$ indicative for a possible crash of the economy. Therefore,
in order to estimate a $t_c$, we shall apply the novel ENLF model
to analyze the evolution of prices in Israel looking for multiple
equilibria (or trajectories). In practice, there are two
strategies for treating the parameters of the ENLF model. One is
to adopt as free parameters $r_0$, $\beta$, $\gamma$, and $t_c$
like it was done for the analysis of data for Hungary. The other,
is to consider $r_0$, $\beta$, $\gamma$, and $z_0$ as free
parameters. It is important to notice that the whole contribution
of $t_c$ is carried by $z_0$ in the form of the product $\beta
\times (t_c-t_0)/\Delta t$, which also determines $a_p({\rm MTT})$
as shown above. Hence, it would be reasonable to expect a solution
compatible with the LF model. Therefore, in order to check this
feature, the CPI data for the period 1969-1985 were, in a first
step, fitted to Eq.\ (\ref{lptng5}) expressed in therms of $z_0$.
The obtained parameters are listed in Table \ref{tab:table4},
where one can observe that the $\chi$ is similar to that obtained
for the long NLF run. The value $\beta=0.022\pm0.117$ is
consistent with zero suggesting a LF with a renormalized strength
given by Eq.\ (\ref{rate48}) with $\beta=0$
\begin{equation}
a_p(\beta \to 0) = a_p({\rm MTT}) - \frac{\gamma -1}{2\,\gamma}
\;. \label{a_p_g}
\end{equation}
In fact, the results for $a_p(z_0)$ [evaluated from Eqs.\
(\ref{singul_1}) and (\ref{defxb0})] and $a_p({\rm MTT})$ quoted
in Table \ref{tab:table4} satisfy this relation. This solution
remains stable when the minimization procedure is continued.
Furthermore, the use of Eq.\ (\ref{defxb0}) with the listed
results for $z_0$, $\beta$, and $\gamma$ yields $t_c=2170$, which
is too large for a hyperinflation developing during the 1980's.
So, the prediction of this solution is not useful.

The fit considering $r_0$, $\beta$, $\gamma$, and $t_c$ as free
parameters yielded the set of results reported in Table
\ref{tab:table4}. The obtained values $t_c=2015$, $\beta=0.17$,
and $\gamma=1.07$ are quite reasonable for a rather slow
hyperinflation. If the fit is continued the parameters do not
change, only the uncertainties diminish as can be seen in Table
\ref{tab:table4}. Although the $\chi$ of this solution for $t_c$
is slightly larger than that determined by using $z_0$, it is
quite acceptable. The good quality of the present ENLF(t$_c$)
description of measured data for both CPI and GRI is shown in
Fig.\ \ref{fig:Israel_h}.

The value of $t_c$ obtained by fitting $\ln P(t)$ with Eq.\
(\ref{lptng5}) is larger than that determined by STZ* from a
phenomenological fit of $P(t)$ to Eq.\ (\ref{price1}). This
feature is due to the fact that, for a regime with inflation,
$P(t)$ presents a steeper slope than $\ln P(t)$
\begin{equation}
\frac{d P(t)}{dt} = P(t)\,\frac{d \ln P(t)}{dt} \;. \label{deP}
\end{equation}
The steeper is the curvature the earlier becomes $t_c$ [see Fig.\
\ref{fig:Israel_h}(a)].

Since the actions for stopping the inflation begun in 1985
\cite{leiderman03}, we also performed an analysis of the reduced
period 1969-1984. The values yielded by the new sequence of fits
are also included in Table \ref{tab:table4}. There one can observe
small changes between the parameters corresponding to both series
of data.

In summary, we can state that in the case of the hyperinflation
occurred in Israel it is possible to predict a reasonable $t_c$ by
applying the ENLF model proposed in the present paper. In fact,
the interaction between $\beta$ and $\gamma$ leads to
multiequilibria phenomena yielding solutions with $\beta=0$ (no
prediction for $t_c$) and $\beta > 0$ ($t_c$ is determined). As
mentioned in Sec.\ \ref{sec:introduction}, the literature on
multiple equilibria in economics is huge \cite{cooper99,%
diamond82,bruno1987,obstfeld96,krugman99,morris00,nadal05}.

\subsection{High-inflation episodes in Mexico and Iceland}
\label{sec:high}

This section is devoted to study regimes of high-inflation that
are not catalog as episodes of hyperinflation. Looking at tables
of inflations one can find several examples of this type. If
evolution of the CPI of such a sort of episodes is described by
Eq.\ (\ref{price1}) one finds similar features to that
encountered in the case of Israel. For illustrating this kind of
behavior we selected the evolution of prices during periods of
high-inflation occurred in Mexico and Iceland.

\begin{figure*}
\includegraphics[width=8.5cm, height=6cm]{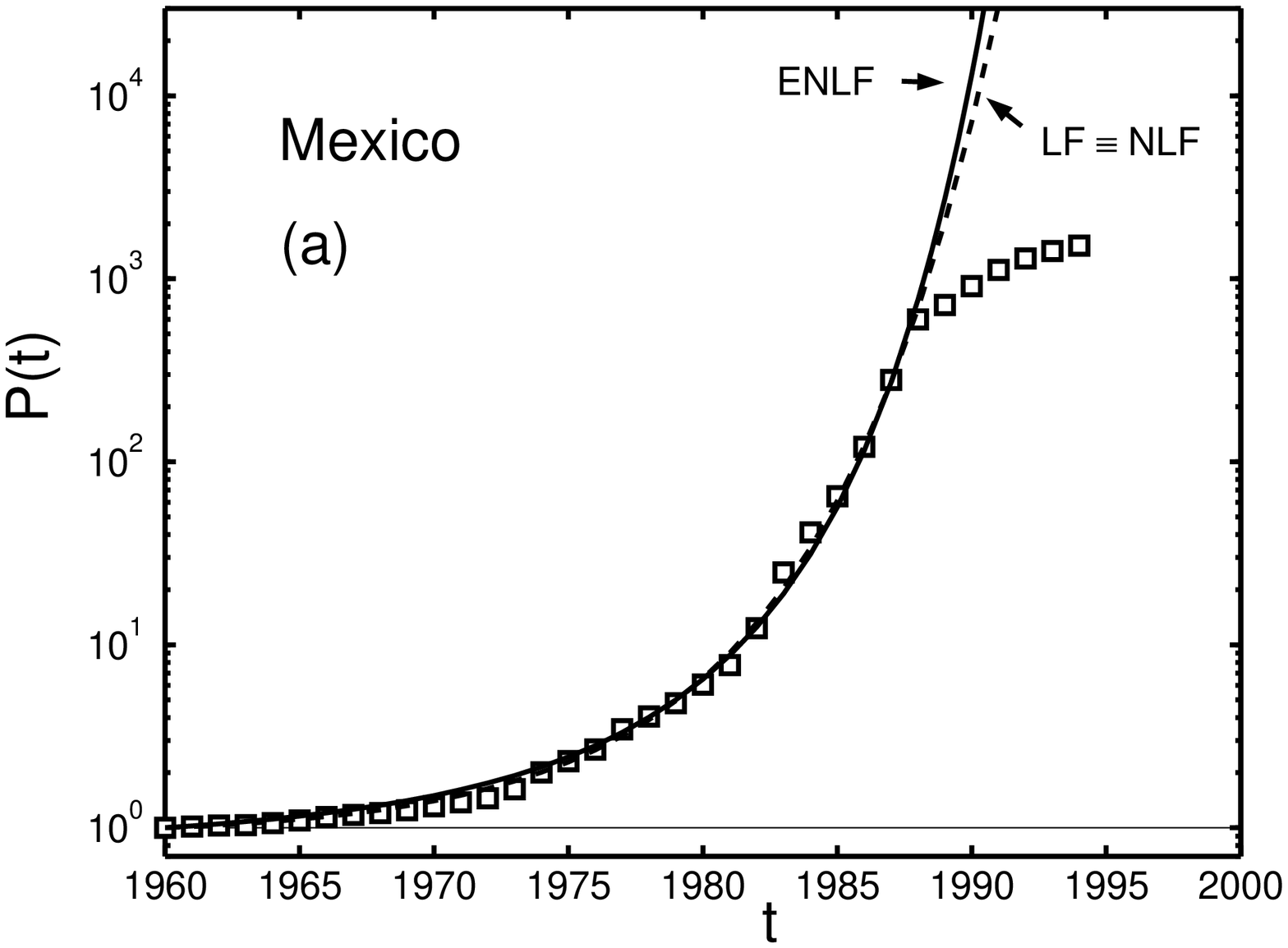}
\includegraphics[width=8.5cm, height=6cm]{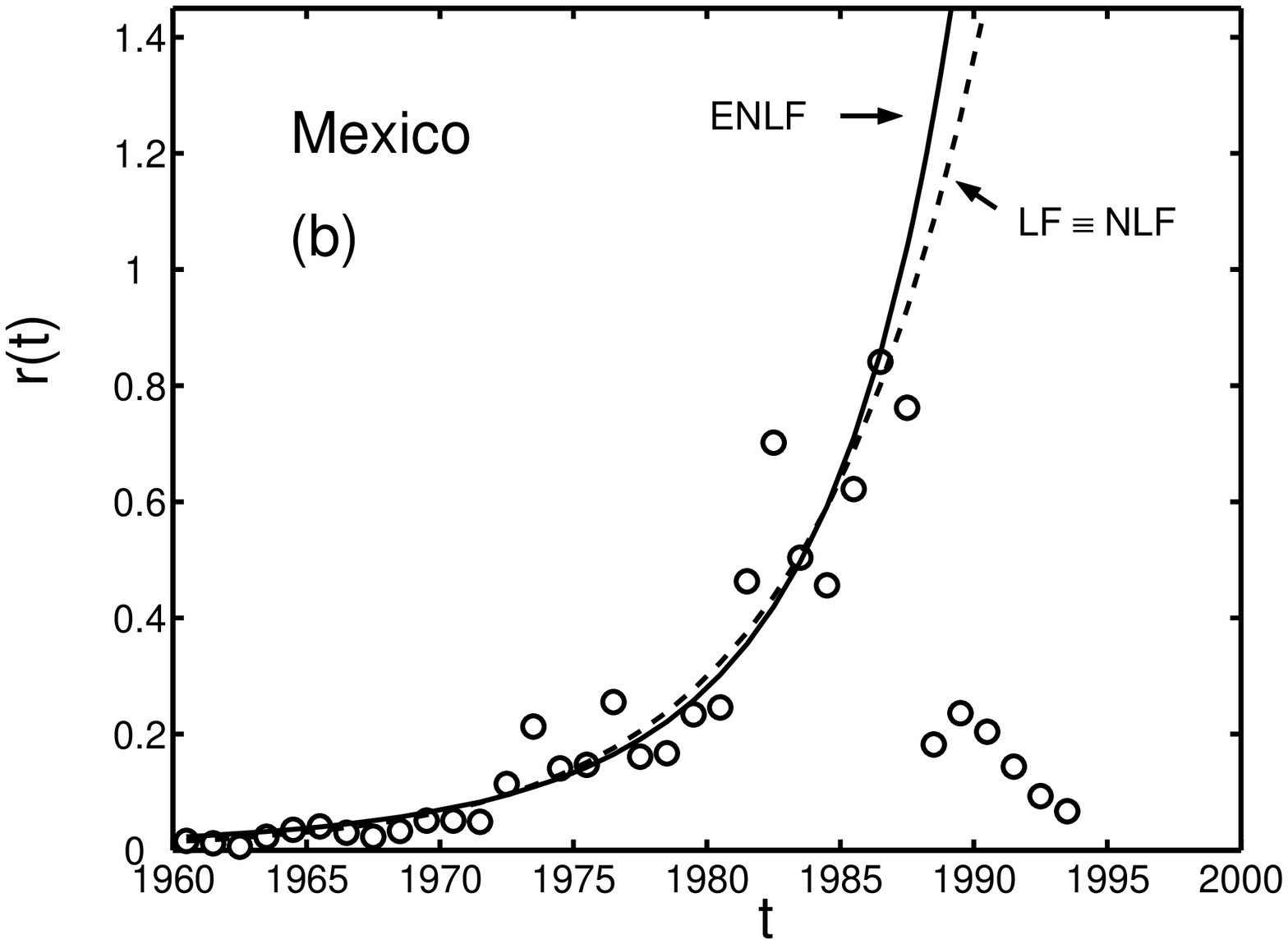}
\caption{\label{fig:Mexico_high} (a) Squares are yearly CPI in
Mexico from 1960 to 1994 normalized to $P(t_0=1960)=1$. The
dashed line is the fit of the series 1960-1988 with Eq.\
(\ref{p_mizu}) of the LF approach, while the solid curve show the
fit of the same data with Eq.\ (\ref{lptng5}) of the present
ENLF($t_c$) model. (b) Circles are yearly GRI for the same period
as in (a). The dashed line was evaluated with Eq.\ (\ref{r_mizu}),
LF model, while the solid curve was calculated with Eq.\
(\ref{rate57}), ENLF($t_c$) model. In both cases the parameters
listed in Table \ref{tab:table4} were used.}
\end{figure*}

The economy of Mexico is rather large, the number of inhabitants
during the period of high-inflation was close to 100 millions. The
general conditions that gave rise to such an inflation have been
based on domestic troubles and the management of petroleum
business \cite{rojas92,goodman97,bergoeing07}. Political unrest
escalated during the 1960's. Students of the Autonomous National
University of Mexico began organizing large scale demonstrations
in Mexico City in 1968. These protests were followed by violent
episodes including the killing of demonstrators by members of the
armed forces \cite{goodman97}. At the beginning of the 70's
president Luis Echeverr\'{\i}a \'Alvarez trying to avoid an
ongoing insurgency distributed land to a communal farming
arrangement. After this step the production decreased
substantially. In addition, he promoted extensive subsidies for
public and private enterprises. For instance, one of the projects
was a steel complex in Michoac\'an. These programs could not be
funded out of existing tax funds so the Central Bank printed
money. One of the effects of these policies was a serious
inflation accompanied by an economic crisis. At the end of the
70's during the presidency of Jos\'e L\'opez Portillo y Pacheco
new oil fields were discovered in the south of Mexico. Because of
the existence of these new reserves and the rising international
price of petroleum, foreign bankers were willing to lend Mexico
vast amount of money. By 1982 almost half of the petroleum exports
earning were going to pay the interest and other scheduled
payments for the foreign debt. So, these discoveries did not
alleviate the problem of inflation, the annual rate of inflation
hit 100\% [see Fig.\ \ref{fig:Mexico_high}(b)]. In September 1982
the banks of Mexico were nationalized. The price of petroleum
began to fall in response to the increased quantity of petroleum
being supplied. When Miguel de la Madrid Hurtado came into office
at the end of 1982 Mexico's economic house was in a great state of
disarray. There was a huge foreign debt requiring excessive
foreign currency credit to service. He promulgated a program of
economic austerity which included: increases in tax rates;
reduction of the federal government budget; reduction of subsidies
for some commodities; postponement of many public projects;
increase of some interest rates; and relaxation of capital
transfer restrictions. The moves toward privatizations started
during his administration. Financial necessity forced the selling
off of about 200 government enterprises. Nevertheless, the
inflation continuously grew, see Fig.\ \ref{fig:Mexico_high}(b).
In 1988, Carlos Salinas de Gotari, young technocrat with a Ph.D.
in economics from Harvard University was elected president. The
criterion was to find a person with expertize for dealing with
Mexico's financial and economic problems. He deepen successfully
the stabilization program for stopping inflation began by de la
Madrid, see Fig.\ \ref{fig:Mexico_high}(b).

\begin{figure*}
\includegraphics[width=8.5cm, height=6cm]{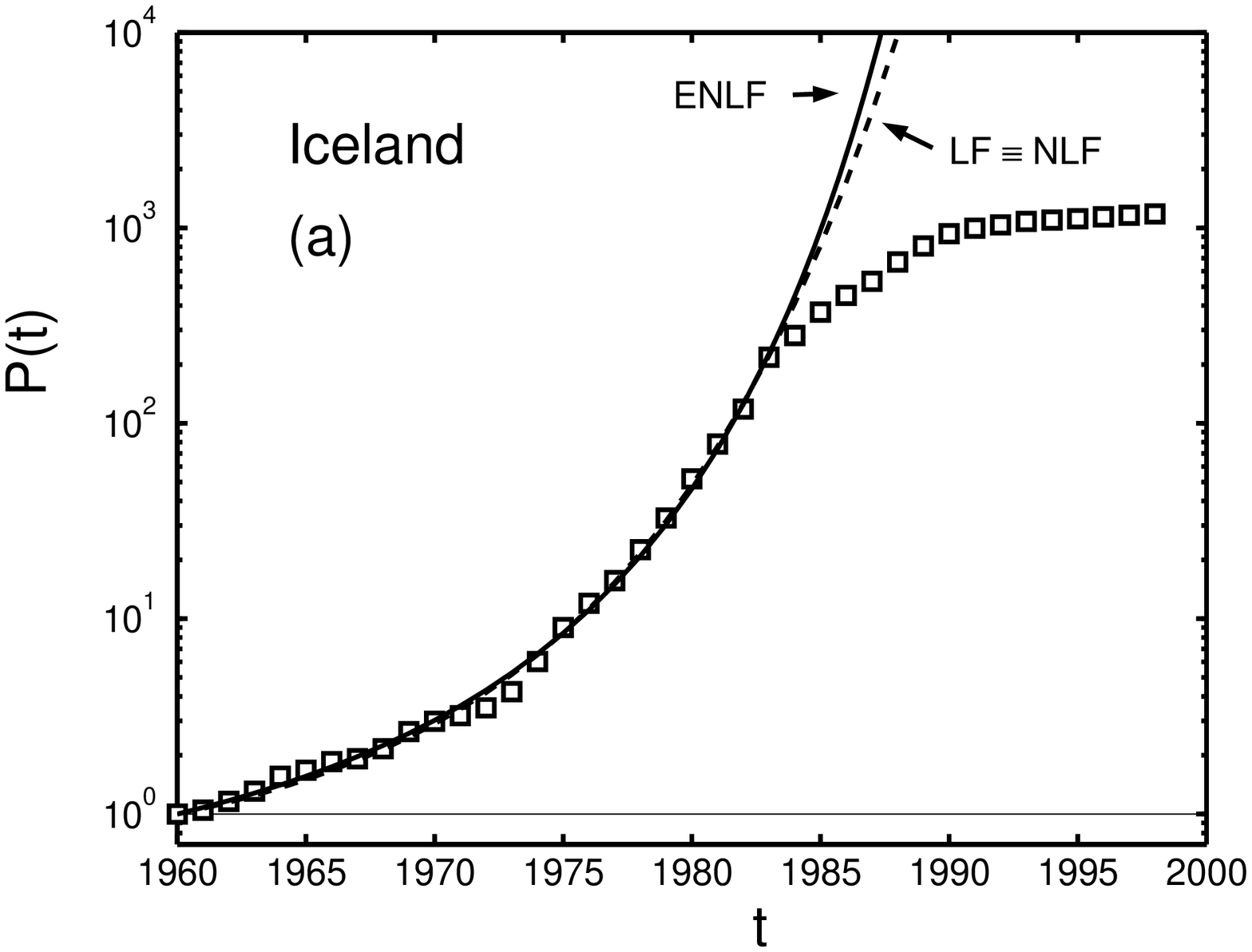}
\includegraphics[width=8.5cm, height=6cm]{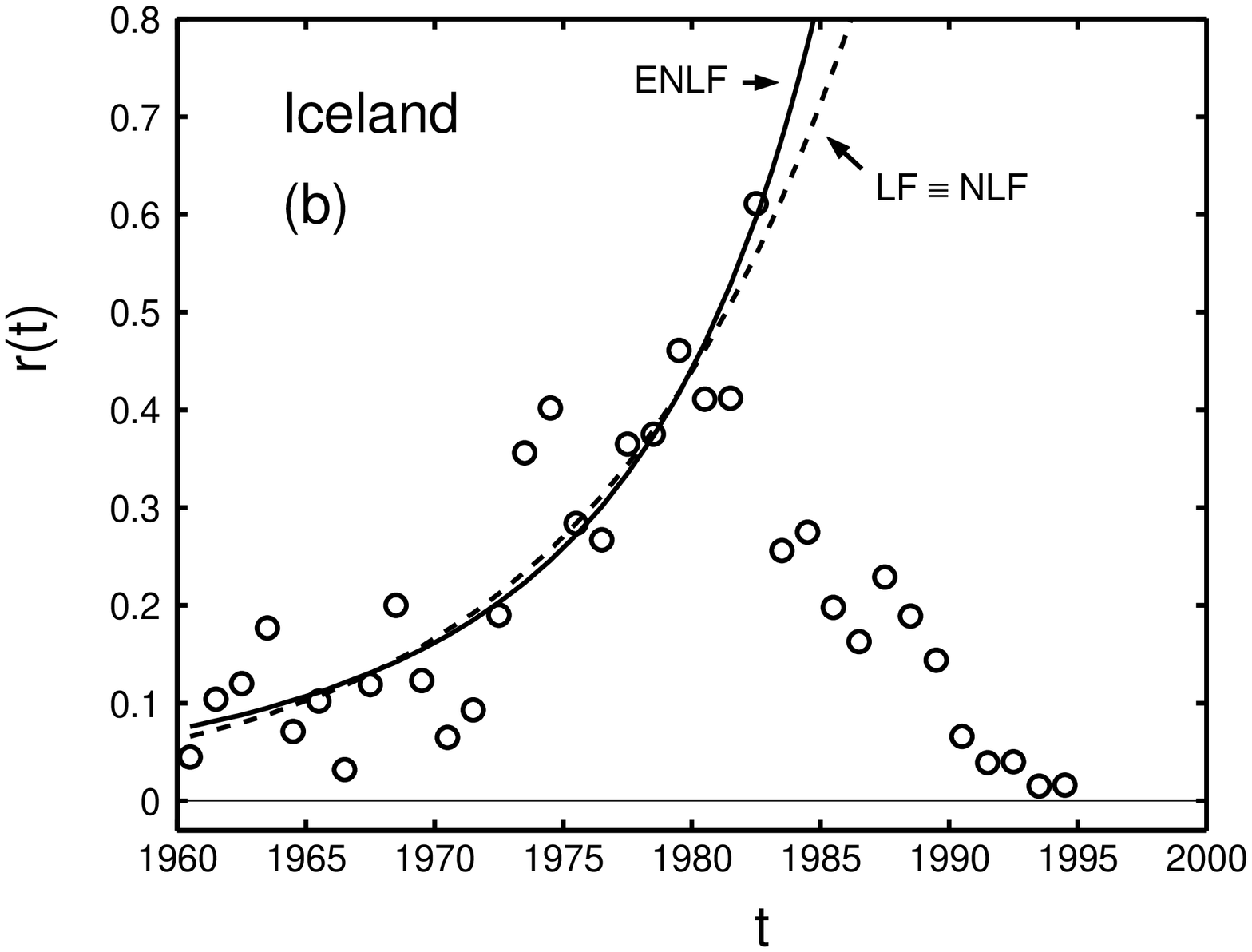}
\caption{\label{fig:Iceland_high} (a) Squares are yearly CPI in
Iceland from 1960 to 1998 normalized to $P(t_0=1960)=1$. The
dashed line is the fit of the series 1960-1983 with Eq.\
(\ref{p_mizu}) of the LF approach, while the solid curve show the
fit of the same data with Eq.\ (\ref{lptng5}) of the present
ENLF($t_c$) model. (b) Circles are yearly GRI for the same period
as in (a). The dashed line was evaluated with Eq.\ (\ref{r_mizu}),
LF model, while the solid curve was calculated with Eq.\
(\ref{rate57}), ENLF($t_c$) model. In both cases the parameters
listed in Table \ref{tab:table4} were used.}
\end{figure*}

The CPI and GRI were evaluated using data of inflation taken from
Ref.\ \cite{FRED}, these observables are plotted in panels (a) and
(b) of Fig.\ \ref{fig:Mexico_high}, respectively. The values from
1960 to 1988, previous to stabilization, have been analyzed. For
the study of the Mexico's inflation the same procedure as that
applied in the case of Israel was adopted. So, in a first step, we
fitted data of CPI with Eq.\ (\ref{price1}). The parameters
together with the $\chi$ obtained by stopping the minimization
procedure when the variation of $\chi$ between the ``$i+1$'' and
``$i$'' iterations was smaller than a standard choice $10^{-1}\,
\%$ are listed in Table \ref{tab:table4}. A glance at this table
shows a small exponent of the power law $\beta=0.043\pm0.028$ and
a large critical time $t_c = 2132 \pm 99$. If one allows the
iterations to continue, the correlation between these parameters
becomes clear. For instance, in Table \ref{tab:table4} we quoted
values obtained when the change of $\chi$  becomes less than
$10^{-3}\,\%$. As in the case of Israel, $\beta$ goes to $0$ and
$t_c$ increases keeping $a_p$ given by Eq.\ (\ref{a_p_lim})
constant. These results indicate that the NLF model tends towards
the LF one. Therefore, the analysis was completed fitting the CPI
data directly with LF's Eq.\ (\ref{p_mizu}). The obtained
parameters are included in Table \ref{tab:table4}. Notice the 
excellent agreement between the values of $r_0$, $a_p$, and $\chi$ 
yielded by the LF approach and those obtained from the ``long''
fit with Eq.\ (\ref{price1}) of the NLF model. The quality of the
fits is depicted in Fig.\ \ref{fig:Mexico_high}(a). In addition,
the theoretical GRI was calculated using Eqs.\ (\ref{r_time}) and
(\ref{r_mizu}) and it is compared to data in Fig.\ 
\ref{fig:Mexico_high}(b). No difference between LF and NLF can be
observed. So, no estimation for $t_c$ can be achieved.

Next, the CPI data for the period 1960-1988 were fitted to Eq.\
(\ref{lptng5}) provided by the ENLF model written in therms of
$r_0$, $\beta$, $\gamma$, and $z_0$. This procedure lead to the
values listed in Table \ref{tab:table4}. The value $\beta=0.082\pm
0.149$, although larger than in the case of Israel, is also
consistent with zero suggesting a LF. Moreover, the result for
$a_p(z_0)$ also satisfies the relation (\ref{a_p_g}).
 
Finally, the same data of CPI were fitted with Eq.\ (\ref{lptng5})
of the ENLF model written in therms of $r_0$, $\beta$, $\gamma$,
and $t_c$. The obtained parameters are also included in Table
\ref{tab:table4} and the fit is displayed in Fig.\
\ref{fig:Mexico_high}(a). In this case, both parameters $t_c=2034$
and $\beta=0.134$ are well defined and reasonable. Although the
$\chi$ is slightly larger than that obtained with the LF model a
good match of theoretical CPI with measured data is got. For
completeness, the GRI was evaluated using Eq.\ (\ref{rate57}) and
plotted in Fig.\ \ref{fig:Mexico_high}(b), where one may observe a
good accordance with data. These results indicate that the
ENLF($t_c$) model provides a satisfactory description of the
episode occurred in Mexico. Hence, one can state that multiple
equilibria phenomena are also present in this case. The solution
with $\beta > 0$ shows a trajectory towards the category of
hyperinflation.

It is also interesting to study the increase of prices occurred in
Iceland during the period from 1960 to 1983 \cite{andersen98}
right previously to the disinflation analyzed by MTT (see Table 2
in Ref.\ \cite{mizuno02}). Let us mention that Iceland is a nation
with less than a half million inhabitants, so in contrast to
Mexico this is a rather small economy.
Iceland was under Norwegian and Danish kings along centuries. It
is an independent nation since 1944. During the examined period
the government's fiscal policy was strictly Keynesian, and their
aim was to create the necessary industrial infrastructure for a
prosperous developed country. It was considered essential to keep
unemployment down to an absolute minimum and to protect the export
of fishing industry through currency manipulation and other means.
Due to the country's dependence both on unreliable fish catches
and foreign demand for fish products, Iceland's economy remained
very unstable well into the 1990's, when the country's economy was
greatly diversified. Iceland then became a member of the European
Economic Area in 1994. Economic stability increased and previously
chronic inflation was drastically reduced.

The GRI and CPI for Iceland were evaluated with data taken from
\cite{FRED} and are plotted in panels (a) and (b) of Fig.\ 
\ref{fig:Iceland_high}, respectively. One may observe that before
stabilization the CPI reaches a value of about $5 \times 10^2$,
being more than one order of magnitude smaller than the value
$10^4$ corresponding to Israel and slightly smaller than that of
Mexico. Moreover, a further check of this relative strength of
inflations can be done comparing Fig.\ \ref{fig:Iceland_high}(b)
with Figs.\ \ref{fig:Israel_h}(b) and \ref{fig:Mexico_high}(b).
Hence, the inflation of Iceland was studied in the same
way as the episodes of that countries. However, in this case the 
analysis was begun by fitting data of CPI for the period 1960-1983 
with Eq.\ (\ref{p_mizu}) of the LF model. The obtained parameters
are listed in Table \ref{tab:table4}. The quality of the
adjustment is shown in Fig.\ \ref{fig:Iceland_high}(a). The
calculated $B_{MTT}=1+2a_p=1.20$ is smaller than the values quoted
in Table 1 of Ref. \cite{mizuno02}, indicating that in this case
the inflation was less severe than for the examples examined
there. Next, a standard ``short'' fit of the same data for CPI
with Eq.\ (\ref{price1}) of the NLF model, similar to that
performed in the cases treated previously, yielded the parameters
quoted in Table \ref{tab:table4}. One may realize that $\beta=
0.160\pm 0.260$ is consistent with zero and $t_c=2037\pm104$ is
rather undetermined. Furthermore, a ``longer'' fit indicates that
$\beta$ goes to 0 and $t_c$ increases presenting an even larger
uncertainty, both these features can be seen in Table
\ref{tab:table4}. On the other hand, $a_p$ given by Eq.\
(\ref{a_p_lim}) converges to a constant  value, which coincides
with that yielded by the fit with the LF model. The GRI evaluated
with Eqs.\ (\ref{r_time}) and (\ref{r_mizu}) is displayed in Fig.\
\ref{fig:Iceland_high}(b), no  difference between the LF and NLF
approaches can be observed on the scale of the drawing. So, as
expected, in this case the NLF model also converges towards the LF
one. Hence, no prediction for $t_c$ could be obtained.

Therefore, in order to get an estimation for $t_c$, we also
applied the ENLF model for analyzing this episode. Firstly, the
same series of CPI data was fitted to Eq.\ (\ref{lptng5}) written
in therms of $r_0$, $\beta$, $\gamma$, and $z_0$. This procedure
lead to the values listed in Table \ref{tab:table4}, where one can
realize that the $\chi$ is equal to that obtained with the LF
model. The value $\beta=0.037 \pm1.664$, although presents a very
large uncertainty, it is consistent with zero suggesting a LF. On
the other hand, the result for $a_p(z_0)$ also satisfies the
relation (\ref{a_p_g}). However, the critical time calculated with
Eq.\ (\ref{defxb0}), $t_c=2291$, is too large making this
prediction useless.
 
Finally, the ENLF model written in therms of $r_0$, $\beta$,
$\gamma$, and $t_c$ was applied for describing the examined data.
The fit with Eq.\ (\ref{lptng5}) yielded the parameters are
included in Table \ref{tab:table4}. In this case both parameters,
$t_c=2049$ and $\beta=0.151$, are reasonable. If the minimization
is continued the values of the parameters remain stable, while
their uncertainties diminish. In addition, the $\chi$ is almost
equal to that provided by the fit to the LF model. As depicted in
Fig.\ \ref{fig:Iceland_high}(a), the matching between theoretical
CPI and measured data is quite good. For completeness, the GRI was 
evaluated using Eq.\ (\ref{rate57}) with the parameters of the
ENLF($t_c$) model. The result is plotted in Fig.\
\ref{fig:Iceland_high}(b), where one may observe a good accordance
with data. So, the ENLF($t_c$) model describes satisfactorily well
the episode occurred in Iceland providing an acceptable prediction
for $t_c$ in case that this high inflation would become a
hyperinflation. This is another example of multiple equilibria.

\section{Summary and Conclusions}
\label{sec:summary}

In the present work we treated regimes of hyper- and
high-inflation in economy. In a previous work \cite{szybisz16} it
has been found that for a weak hyperinflation, like e.g. that
developed in Israel, was impossible to determine a value of $t_c$
within the frame of the NLF model. This model is based on a power
law with an exponent $\beta>0$, see Eq.\ (\ref{rate00}). The
mentioned drawback has been attributed to a permanent but
incomplete effort for stopping inflation. Therefore, in the
present work we suggested to include in the theory information on
saturation by introducing a parameter $\gamma$, which multiplies
all the past inflation growth rates. This parameter would account
for the effort done by the government for coping inflation.

In the extended approach, ENLF, reported in the present paper the
solutions for GRI and CPI are also analytic as in the NLF model.
In particular, the CPI is expressed in therms of the Gaussian
hypergeometric function $_2F_1(1/\beta,1/\beta,1+1/\beta;z)$,
where $z$ is a function of $\gamma$, $\beta$, $r_0$, and $t_c$,
see Eqs.\ (\ref{defxb0})-(\ref{defxb}). It is pertinent to notice
that $_2F_1$ appears in a variety of mathematical and physical
problems. For $z \to 1$ this hypergeometric function diverges
leading to a finite time singularity, from which a value of $t_c$
can be determined. The same singularity is present in Eq.\
(\ref{rate57}) for GRI. So, the ENLF model proposed in the present
work preserves the well-defined singularities yielded by the power
law stemmed from a simple positive nonlinear feedback. This
mechanism is important for understanding processes in financial
crashes (see Ref.\ \cite{sornette03b} and references therein). For
completeness, in the appendix it is shown that for the limit
$\gamma \to 1$ from above, one retrieves all the expressions of
the NLF model.

An analysis of the severe hyperinflation occurred in Hungary after
the World War II proves that the novel ENLF approach is robust.
When it is used for examining data of Israel there are two sorts
of solutions. One yields $\beta$ consistent with zero (LF model)
and the other one gives a well determined and reasonable $t_c$.
As a further application, high-inflation regimes exhibiting weaker
inflations than that of Israel were also analyzed. The episodes
occurred in Mexico and Iceland are reported in the present work.
Data of both series of inflation can be described, as in the case
of Israel, with the LF model. However, the ENLF($t_c$) model also
provides additional solutions forecasting possible blow up of the
economies in case the high inflation regimes would become spirals
of hyperinflation. The corresponding fits are very good and the
predicted values of $t_c$ for crashes are acceptable.

The phenomena of multiple equilibria, a known feature in models of
economics \cite{cooper99,diamond82,bruno1987,obstfeld96,%
krugman99,morris00,nadal05}, appears due to the fact that the
introduction of $\gamma$ enlarges the dimension of the $\chi$
hyper-surface, which now also presents minimums in domains where
the parameters $t_c$ and $\beta$ are not strongly correlated. So,
we can state that the parameter $\gamma$ allows a more complete
representation of inflationary processes. It is interesting to
note that the exponent of the nonlinear feedback of inflation can
be controlled by a parameter multiplying the growth rate.
Different combinations of parameters may be interpreted as forces
acting with its own strength, producing dynamical paths that lead
to very different outcomes.

\begin{acknowledgments}
This work was supported in part by the Ministry of Science and
Technology of Argentina through Grants PIP 0546/09 from CONICET
and PICT 2011/01217 from ANPCYT, and Grant UBACYT 01/K156 from
University of Buenos Aires.
\end{acknowledgments}

\appendix
\section*{Appendix: Limits for $\gamma \to 1$ ($s \to 0$)}
\label{sto0}

In this appendix we show that by imposing the limit $\gamma \to
1$, which is equivalent to $s \to 0$, in the generalized
expressions for GRI and CPI the forms reported in Ref.\
\cite{szybisz09} are recovered. So, starting from Eq.
(\ref{rate55}) for GRI and keeping only linear terms of the
expansion in`powers of $s$ one gets
\begin{eqnarray}
r(t) &=& r_0 \biggr[\frac{s\,\exp[\beta\,s\,(t-t_0)]}{s + q\,
r_0^\beta ( 1 - \exp[\beta\,s\,(t-t_0)]) } \biggr]^{1/\beta}
\nonumber\\
&=& r_0 \biggr[\frac{s\,[1-\beta\,s\,(t-t_0)]}{(s + q\,r_0^\beta )
\,[1-\beta\,s\,(t-t_0)] - q\,r_0^\beta}\,\biggr]^{1/\beta}
\;, \label{B_1}
\end{eqnarray}
which in the limit $s \to 0$, where $q \to 1/[\beta r^\beta_0
(t_c-t_0)]$, yields
\begin{eqnarray}
r(t) &=& r_0 \biggr[\frac{1}{1-(t-t_0)/(t_c-t_0)}\biggr]^{1/\beta}
\nonumber\\
&=& r_0 \biggr(\frac{t_c-t_0}{t_c-t}\biggr)^{1/\beta} \;,
\label{B_2}
\end{eqnarray} 
recovering Eq. (\ref{r_time}).

(i) Starting from the general solution for CPI given by Eq.\
(\ref{lptng3}) written as
\begin{eqnarray}
\ln \left[\frac{P(t)}{P_0}\right] &=& \frac{r_0}{\Delta t}
\biggr( \frac{1}{s + q r_0^\beta} \biggr)^{1/\beta}\,s^{-1+
1/\beta} \nonumber\\
& \times & \biggr[\, _2F_1(1/\beta,1/\beta,1+1/\beta;z)
\, \exp[s(t-t_0)] \nonumber\\
&& - \, _2F_1(1/\beta,1/\beta,1+1/\beta;z_0) \biggr] \;,
\label{B_3}
\end{eqnarray}
the limit $s \to 0$ is evaluated expanding the hypergeometric
function for small values of $s$. Using the relation
\begin{eqnarray}
&&z\,\frac{d}{dz}\biggr[\, _2F_1(1/\beta,1/\beta,1+1/\beta;z)
\biggr] = (1-z)^{-1/\beta} \nonumber\\
&&-\, _2F_1(1/\beta,1/\beta,1+1/\beta;z) \biggr] \;, \label{deriv}
\end{eqnarray}
and the on line Mathematica one gets
\begin{eqnarray}
&&s^{-1+1/\beta}\, _2F_1(\frac{1}{\beta},\frac{1}{\beta},1+
\frac{1}{\beta};z) = \nonumber\\
&& s^{-1+1/\beta}\, _2F_1(\frac{1}{\beta},\frac{1}{\beta},1+
\frac{1}{\beta};z) \biggr|_{s=0} \nonumber\\
&&~~~~~~ \times \biggr[ 1 + \biggr( \frac{1}{\beta\,q\,r^\beta_0}
- (t-t_0) \biggr)\,s \biggr] \nonumber\\
&& + \, \frac{\Gamma[1+\frac{1}{\beta}] \Gamma[-1+\frac{1}
{\beta}]}{(\Gamma[\frac{1}{\beta}])^2} \biggr( \frac{1}{q\,
r^\beta_0} - \beta\,(t-t_0) \biggr)^{1-1/\beta} \nonumber\\
&& + \, {\cal{O}}(higher-order) \;. \label{B_4}
\end{eqnarray}
Upon introducing this result into Eq. (\ref{B_3}) and using the
property of $\Gamma$ functions
\begin{equation}
\frac{\Gamma[1+1/\beta] \Gamma[-1+1/\beta]}{(\Gamma[1/\beta])^2}
= \frac{1}{1-\beta} \;, \label{gammas}
\end{equation}
the expansion of the exponential yields
\begin{eqnarray}
&& \ln \left[\frac{P(t)}{P_0}\right] = \frac{r_0}{\Delta t}
\biggr( \frac{1}{s + q r^\beta_0} \biggr)^{1/\beta} \biggr\{
\biggr\{ s^{-1+1/\beta} \nonumber\\
&& \times\, _2F_1(\frac{1}{\beta},\frac{1}{\beta},1+\frac{1}
{\beta};z) \biggr|_{s=0} \biggr[ 1 + \biggr( \frac{1}{\beta q
r^\beta_0} - (t-t_0) \biggr) s \biggr] \nonumber\\
&& + \frac{1}{1-\beta} \biggr( \frac{1}{q\,r^\beta_0}
- \beta\,(t-t_0) \biggr)^{1-1/\beta} \biggr\} [1+(t-t_0)\,s]
\nonumber\\
&& - s^{-1+1/\beta}\, _2F_1(\frac{1}{\beta},\frac{1}{\beta},
1+\frac{1}{\beta};z_0) \biggr|_{s=0}
\biggr[ 1 + \frac{s}{\beta q r^\beta_0} \biggr] \nonumber\\
&& - \frac{1}{1-\beta} \biggr( \frac{1}{q\,r^\beta_0}
\biggr)^{1-1/\beta} \biggr\} \;, \label{B_5}
\end{eqnarray}
When keeping in the CPI only the lowest order of $s$ one gets
\begin{eqnarray}
&& \ln \left[\frac{P(t)}{P_0}\right] = \frac{r_0}{\Delta t}
\biggr( \frac{1}{s + q r^\beta_0} \biggr)^{1/\beta} \biggr\{
s^{-1+1/\beta} \nonumber\\
&& \times\, _2F_1(\frac{1}{\beta},\frac{1}{\beta},1+\frac{1}
{\beta};z) \biggr|_{s=0} \biggr[ 1 + \frac{s}{\beta q
r^\beta_0} \biggr] \nonumber\\
&& + \frac{1}{1-\beta} \biggr( \frac{1}{q\,r^\beta_0}
- \beta\,(t-t_0) \biggr)^{1-1/\beta} [1+(t-t_0)\,s]
\nonumber\\
&& - s^{-1+1/\beta}\, _2F_1(\frac{1}{\beta},\frac{1}{\beta},
1+\frac{1}{\beta};z_0) \biggr|_{s=0}
\biggr[ 1 + \frac{s}{\beta q r^\beta_0} \biggr] \nonumber\\
&& - \frac{1}{1-\beta} \biggr( \frac{1}{q\,r^\beta_0}
\biggr)^{1-1/\beta} \biggr\} \;. \label{B_6}
\end{eqnarray}
Furthermore, due to the fact that
\begin{equation}
\, _2F_1(\frac{1}{\beta},\frac{1}{\beta},1+\frac{1}{\beta};z)
\biggr|_{s=0} = \, _2F_1(\frac{1}{\beta},\frac{1}{\beta},
1+\frac{1}{\beta};z_0) \biggr|_{s=0} \:, \label{F_limit}
\end{equation}
the remaining contributions of the Gauss' hypergeometric
function cancel out leading to
\begin{eqnarray}
&& \ln \left[\frac{P(t)}{P_0}\right] = \frac{r_0}{\Delta t}
\biggr( \frac{1}{s + q r_0^\beta} \biggr)^{1/\beta}
\frac{1}{1-\beta} \biggr( \frac{1}{q\,r^\beta_0}
\biggr)^{1-1/\beta} \nonumber\\
&& \times \biggr[ \biggr( 1 - \beta\,q\,r^\beta_0\,(t-t_0)
\biggr)^{1-1/\beta} [1+(t-t_0)\,s] - 1 \biggr] \;. \nonumber\\
\label{B_7}
\end{eqnarray}
Now, the limit $s \to 0$, where $q \to 1/[\beta r^\beta_0
(t_c-t_0)]$, leads to
\begin{eqnarray}
\ln \left[\frac{P(t)}{P_0}\right] &=& \frac{r_0}{(1-\beta)
\Delta t} \frac{1}{q r^\beta_0} \nonumber\\
&&~~ \times \biggr[ \biggr(\frac{1}{1-\beta q r^\beta_0 (t-t_0)}
\biggr)^{\frac{1-\beta}{\beta}} - 1 \biggr] \nonumber\\
&=& \frac{\beta\,\,r_0}{(1-\beta)} \biggr(\frac{t_c-t_0}{\Delta t} 
\biggr)\biggr[ \biggr(\frac{t_c-t_0}{t_c-t} \biggr)^{\frac{1-
\beta}{\beta}} - 1 \biggr] \;, \nonumber\\
\label{p_final}
\end{eqnarray}
in agreement with Eq. (\ref{price1}).

(ii) In the case of $\beta=1$, starting from the CPI given by
Eq. (\ref{lptln1})
\begin{equation}
\ln \left[\frac{P(t)}{P_0}\right] = \frac{1}{q\,\Delta t}
\ln \biggr[ \frac{s}{s+q\,r_0\,(1-\exp[ s (t-t_0)])} \biggr] \;,
\label{B_15}
\end{equation}
for $s \to 0$ one gets
\begin{eqnarray}
\ln \left[\frac{P(t)}{P_0}\right] &=& \frac{1}{q\,\Delta t}
\ln \biggr[ \frac{1}{1-q\,r_0\,(t-t_0)} \biggr] \nonumber\\
&=& r_0 \biggr(\frac{t_c-t_0}{\Delta t}\biggr)
\ln \biggr[ \frac{1}{1-(t-t_0)/(t_c-t_0)} \biggr] \nonumber\\
&=& r_0 \biggr(\frac{t_c-t_0}{\Delta t}\biggr)
\ln \biggr( \frac{t_c-t_0}{t_c-t} \biggr) \;, \label{B_16}
\end{eqnarray}
which coincides with Eq. (\ref{price2}).

\widetext

\end{document}